\documentclass[a4paper,11pt]{article}

\usepackage{jheppub}

\usepackage{color}
\usepackage{tabularx}
\usepackage{slashed}

\usepackage{hyperref}
\usepackage{xspace}

\usepackage{amsmath}
\usepackage{amssymb}
\usepackage{graphicx,multirow,subfigure}
\usepackage{float}
\usepackage[T1]{fontenc}

\usepackage[normalem]{ulem}

\def \be{\begin{equation}}
\def \ee{\end{equation}}
\newcommand{\bea}{\begin{eqnarray}}
\newcommand{\eea}{\end{eqnarray}}
\def \nn{\nonumber}

\newcommand{\Vcb}{\ensuremath{|V_{cb}|}\xspace}
\newcommand{\BD}{\ensuremath{\bar B\to D\ell\bar\nu}\xspace}
\newcommand{\BRBzeroD}{\ensuremath{\mathrm{BR}(\bar B^0\to D^+\ell\bar\nu)}\xspace}
\newcommand{\BRBminusD}{\ensuremath{\mathrm{BR}(B^-\to D^0\ell\bar\nu)}\xspace}
\newcommand{\BDs}{\ensuremath{\bar B\to D^*\ell\bar\nu}\xspace}
\newcommand{\BDDs}{\ensuremath{\bar B\to D^{(*)}\ell\bar \nu}\xspace}

\newcommand{\BX}{\ensuremath{\bar B\to X\ell\bar\nu}\xspace}
\newcommand{\Rpmz}{\ensuremath{R^{\pm0}}\xspace}

\definecolor{darkgreen}{rgb}{0.1,0.6,0.3}

\makeatletter
\g@addto@macro\bfseries{\boldmath}
\let\Hy@backout\@gobble
\makeatother

\title{{\boldmath $\bar B\to D^{(*)}\ell\bar \nu$} Branching Ratios and Evidence for Isospin Breaking in {\boldmath $\Upsilon(4S)$} Decays}

\author[a]{Martin Jung}
\emailAdd{martin.jung@unito.it}
\affiliation[a]{Dipartimento di Fisica, Università di Torino \& INFN, Sezione di Torino, I-10125~Torino, Italy}

\author[b]{Stefan Schacht}
\emailAdd{stefan.schacht@durham.ac.uk}
\affiliation[b]{Institute for Particle Physics Phenomenology, Department of Physics, Durham University, Durham DH1 3LE, United Kingdom}

\preprint{
\begin{minipage}{.5\textwidth}
    \raggedleft 
    IPPP/26/30
\end{minipage}
}

\abstract{
We introduce a new method for the determination of the ratio of production fractions $R^{\pm0}=\mathcal B(\Upsilon(4S)\to B^+B^-)/\mathcal B(\Upsilon(4S)\to B^0\bar B^0)$ based on $\bar B\to D^{(*)}\ell\bar \nu$ decays.
Given the importance of these modes, we perform a comprehensive analysis of the available data, extracting the information on their branching fractions and $R^{\pm0}$ in parallel and providing their correlations in order to avoid double-use of this information in phenomenological analyses. We obtain the most precise value for $R^{\pm0}$ from a single channel so far, about 2$\sigma$ from unity. The combination with previously available determinations from other channels yields $R^{\pm0}=1.062(19)$, constituting evidence for isospin violation in $\Upsilon(4S)$ decays.  This demonstrates the necessity to take this effect into account in experimental and phenomenological analyses. The results for the $\bar B\to D^{(*)}\ell\bar \nu$ branching fractions are up to $1.6\sigma$ larger compared to averages available in the literature, owing to the removal of overlooked inconsistencies in the treatment of older analyses and correcting for d'Agostini bias where possible, thereby reducing the $V_{cb}$ puzzle.
}

\begin{document}

\maketitle

\section{Introduction}
\label{sec:intro}

Measurements of absolute branching ratios (BRs) of $B$ mesons are an essential part of flavour physics. They play a key role in the determination of the Cabbibo-Kobayashi-Maskawa~(CKM) matrix elements, both by providing direct access to the $V_{cb}$ and $V_{ub}$ elements and as ingredients to symmetry analyses used to determine the angles of the unitarity triangle. The CKM matrix elements in turn are not only fundamental parameters of the Standard Model (SM) of particle physics, but also form the basis for countless searches for physics beyond the SM (BSM). 

The sheer amount of $B$ mesons produced at modern experiments focused on $B$ physics like Belle~II and LHCb has led to a situation in which many of the BR measurements are already dominated by their systematic uncertainties, and in the future this will be the case for many more measurements. In particular, systematic uncertainties related to the production of $B$ mesons have become an obstacle to improve the overall precision of BRs of $B$ decays and thereby of the underlying physics parameters. The key difficulty is separating the production from the decay process, since generally experimentally measured quantities are only sensitive to their product. 

The production of charged and neutral $B$ mesons at the $B$ factories is of particular importance in this context: this is because it affects almost all BR measurements performed there, and by extension virtually all BR measurements in general, since the measurements at hadron colliders rely on normalization modes obtained from the $B$ factories. It turns out that determining these production fractions is highly non-trivial, see recent discussions in Refs.~\cite{HeavyFlavorAveragingGroupHFLAV:2024ctg,Jung:2015yma,Bernlochner:2023bad}. On the one side, as emphasized in these references, it is desirable to have measurements of the production fractions that are independent of specific $B$ decay modes; this is since otherwise circular arguments might arise regarding isospin violation in the modes used to obtain the production fractions, like $B\to J/\psi K$. On the other hand, the presently available information is limited, and it is important to obtain as precise values for the production fractions of charged and neutral $B$ mesons at the $B$ factories as possible. It is particularly interesting to establish whether 
these fractions show indeed a significant asymmetry,
as hinted at by present averages, albeit so far only at the level of $\sim~2-2.5\sigma$~\cite{HeavyFlavorAveragingGroupHFLAV:2024ctg,Jung:2015yma,Bernlochner:2023bad}. To that aim, we consider in this work the decay modes \BDDs, which are expected to show particularly small isospin violation 
as the spectator-quark dependence is additionally suppressed due to heavy-quark symmetry \cite{Voloshin:1998nh,Kobach:2019kfb},
and are thereby suited to obtain information on $B$ production.  
However, the importance of these modes lies predominantly in the determination of \Vcb, and are under particular scrutiny given the so-called \emph{$V_{cb}$ puzzle}, a significant tension between the \Vcb values extracted from inclusive and exclusive decays, see, for instance, Refs. \cite{Voloshin:1997zi, HeavyFlavorAveragingGroupHFLAV:2024ctg,ParticleDataGroup:2024cfk,Gambino:2019sif, Martinelli:2023fwm,  Bordone:2024weh, Bernlochner:2022ywh, Fedele:2023ewe, Ray:2023xjn,FermilabLattice:2021cdg, Harrison:2023dzh, Aoki:2023qpa, Bordone:2025jur, Bordone:2021oof, Finauri:2023kte, Fael:2024fkt}. Given this situation,
the purpose here is two-fold: on the one hand, to improve the determination of the $B$ production fractions at the $B$ factories, for which these modes  have not been used so far; on the other hand, to obtain a more accurate and reliable determination of the \BDDs BRs themselves, in order to improve on the determination of \Vcb from these modes in a follow-up analysis \cite{Gambino:inpreparation}. 
Regarding the latter aspect, 
the dominance of systematic uncertainties in most measurements renders the inclusion of correlations due to external inputs mandatory.
While this part is similar to the HFLAV analysis \cite{HeavyFlavorAveragingGroupHFLAV:2024ctg}, ours differs in a number of points:
\begin{enumerate}
\item We account for d'Agostini bias where possible, which has been shown to have a significant impact on these modes in Refs.~\cite{Jung:2018lfu,Gambino:2019sif}, see also the discussion below.
\item When using external inputs for the $B$ production fractions, we employ the full analysis from Ref.~\cite{Bernlochner:2023bad}, which already improved on earlier determinations.
\item We consider \BD and \BDs decays simultaneously, 
in order to obtain the correlations between their BRs, which affects the value of \Vcb obtained in a global analysis. 
\item We correct for inconsistencies in the treatment of older analyses that have been overlooked so far.
\end{enumerate}
As a result we obtain together with information on $B$-meson production at the $B$ factories four correlated BRs for \BDDs that can be used as inputs in analyses of these modes within or beyond the SM.

The article proceeds as follows: 
In Sec.~\ref{sec:branchingratios} we discuss the general structure of branching-ratio measurements with a particular emphasis on relevant isospin-breaking effects. In Sec.~\ref{sec:results}, after describing our treatment of common input parameters and correlations, we analyze the available experimental information and present our results for the branching-ratio averages and production fractions. We conclude in Sec.~\ref{sec:conclusions}. Additional details on the individual measurements entering our analysis are provided in the Appendix.

\section{Branching ratio measurements and isospin \label{sec:branchingratios}}

Measurements of absolute BRs are generally obtained by counting events corresponding 
to the decay of a $B$ meson $B^{\pm,0}$ into a corresponding final state $X$, which potentially decays further into the detected final-state $Y$, with the generic structure
\begin{align}\label{eq::Nexp}
    &N(B^{\pm,0}\to X(\to Y) ) =
    \epsilon_{B^{\pm,0}\to X(\to Y)} N_{B^{\pm, 0}} BR(B^{\pm,0}\to X) BR(X\to Y)\,,
\end{align}
where $\epsilon$ denotes the overall experimental efficiency for this decay chain, and $N_{B^{\pm,0}}$ is the number of initial $B^{\pm,0}$ mesons, which depends on the experiment and the details of the measurement: for instance, at the $B$ factories, this quantity is often expressed as 
\begin{align}
    N_{B^{\pm,0}}^{\text{B-factory}}=2N_{\Upsilon(4S)} f_{\pm,00}\,,
\end{align}
with the number of $\Upsilon(4S)$ determined in a separate measurement and the production fractions $f_{\pm,00}$ subject of the present analysis. The main difficulty, alluded to in the introduction, is the separation of the different terms in these equations. For the purpose of this article, we consider the secondary decay as well as the experimental efficiency and the number of $\Upsilon(4S)$ mesons as external inputs. Clearly, in order to obtain a precision determination of the quantities of interest, all of these external inputs are required to meet the same level of precision as the quantities discussed in the following, which is challenging in itself.

Assuming the other ingredients are provided as external inputs, the remaining task is to separate the production fractions from the main branching fractions themselves. This is non-trivial, because none of the two factors in the product are a priori known independently. A few methods to achieve final-state-independent determinations of the production fractions have been discussed in Refs.~\cite{MARK-III:1985hbd,BaBar:2005uwr,Gronau:2006ei,Jung:2015yma,Bernlochner:2023bad}, and, as emphasized above, it is still highly desirable to obtain a high-precision determination of the production fractions this way. However, since such determinations
do not exist yet, we investigate in the following the use of \BDDs decays as a new method to obtain information on the ratio of production fractions
\begin{align}
    R^{\pm 0} \equiv \frac{
        \Gamma(\Upsilon(4S)\to B^+B^-))
        }{
        \Gamma(\Upsilon(4S)\to \bar B^0B^0))
        } = \frac{f_{\pm}}{f_{00}}\,.
\end{align}
This ratio has played the major role in the discussion of production fractions in the past for various reasons:
\begin{enumerate}
    \item It is the central quantity to relate a measured asymmetry between charged- and neutral-$B$ events with the underlying isospin asymmetry in their decay. Despite being predicted itself to be unity in the isospin limit, large deviations from this limit can be expected due to the vicinity of the $B\bar B$ threshold \cite{Atwood:1989em}, so assuming \Rpmz to be unity is \emph{not} justified.  
        \item It is easier to determine than the absolute decay rates: importantly, at the $B$ factories the number of $\Upsilon(4S)$ cancels, and for its determination it is not necessary to know the fraction of $\Upsilon(4S)$ decays to final states other than two $B$ mesons, $f_{\not B}$. Furthermore, considering the isospin limit for the $B$ decay, it is directly accessible from the ratio of counting rates for two isospin-related modes. 
    \item If $f_{\not B}$ is assumed to vanish, as was commonly done in the past, this ratio determines \emph{both} fractions $f_{\pm,00}$, since they fulfill in that limit the additional relation $f_{\pm}~+~f_{00}~=~1$.
\end{enumerate}
While the latter assumption should not be made in a precision determination of production fractions, the other two advantages remain. Assuming again the other ingredients like the efficiency ratio to be provided externally, the theoretical limitation is then given by the precision of the assumption of isospin symmetry, which we discuss in the following.

\subsection{Isospin symmetry in $B$ production}

The production processes for $B$ mesons are very different for the different experiments that provide information on \BDDs decays. We briefly describe the role of isospin symmetry in the corresponding production processes:
\begin{itemize}
    \item $B$ factories: as mentioned above, while formally $\Rpmz=1$ in the isospin limit, this is not expected to be a good symmetry due to the nearby $B\bar B$ threshold. Already the associated phase-space factor yields a symmetry breaking of $4.7(8)\%$.
    A puzzling observation is that further enhancement of the symmetry breaking, as estimated in Refs.~\cite{Atwood:1989em, Lepage:1990hm, Byers:1990rd, Kaiser:2002bm, Voloshin:2003gm, Voloshin:2004nu, Dubynskiy:2007xw, Milstein:2021fnc}, seems to be small or absent \cite{Bernlochner:2023bad}. Nevertheless, the isospin limit for $B$ production at the $B$ factories cannot be considered a sound assumption and we do not use it in our work.
    \item LHC experiments: the situation at the LHC is very different, but similarly complicated. The proton-proton initial state is clearly \emph{not} an isospin singlet and simple expectations for $B$ production cannot be inferred from that. The reason why the production fractions for charged and neutral $B$ mesons are nevertheless expected to be similar is that the main processes for $B$ production are isospin symmetric. However, given the complex environment and the difficulty to quantify this expectation, this assumption should be experimentally verified. 
    An analysis of $B\to \bar DD$ decays showed a $\sim 2\sigma$ indication of a non-zero asymmetry~\cite{Davies:2023arm}, while recent measurements at higher $p_T$ values 
    do not indicate a large asymmetry in $B$ production \cite{CMS:2022wkk,CMS:2025zle}.
    In any case, our analysis does not involve measurements that require such an assumption.
    \item LEP: In contrast to the $B$-meson production at the $B$ factories, production via the $Z$ resonance at LEP is not expected to show an enhanced symmetry breaking, since the corresponding threshold is very far away. As a consequence, all $b$-hadron species are produced and tag information can be used to select $b$-hadron events, but not to infer the $B$-meson momentum like at the $B$ factories. In all LEP measurements the yields are proportional to $R_b$, the fraction of $b\bar b$ events in hadronic $Z$ decays, and the fraction of $b$ quarks hadronizing into a neutral $B$ meson, $f_{B^0}$, characterizing the $B$ production. While it should be noted that the extraction of the production fraction for $B$ mesons given by the LEP experiment uses the assumption of equal production of charged and neutral $B$ mesons, the uncertainty of that production fraction is significantly larger than the estimated uncertainty related to isospin-symmetry breaking. We therefore 
    do not assign an additional uncertainty to this assumption in our analysis, 
    keeping however in mind that potential future experiments at the $Z$ pole will need to determine the symmetry breaking in production precisely if they want to achieve their ambitious precision goals.
\end{itemize}

\subsection{Isospin symmetry in \BDDs decays \label{sec:isospin-B-decays}}

In general, the fact that QCD does not distinguish between up and down quarks in the limit $m_u=m_d$ allows to relate matrix elements that involve particles from the same multiplets in the initial and final states. In the simplest cases these relations amount to equalities of matrix elements in the isospin limit. This is the case for $\Upsilon(4S)\to B\bar B$, albeit with large corrections to this limit, but also for \BDDs transitions: both $\bar B$ and $D^{(*)}$ transform as doublets under isospin, while the relevant weak Hamiltonian as well as the involved leptons transform as singlets, hence isospin symmetry implies the trivial relation 
\begin{align}\label{eq::isospin}
    \Gamma(\bar B^0\to D^{(*)+}\ell\bar \nu)\stackrel{\mathrm{isospin}}{=}\Gamma(B^-\to D^{(*)0}\ell\bar \nu)
\end{align}
for $D,D^*$ separately. Corrections to this statement stem from two sources: the first is $m_u\neq m_d$, expected in general to yield corrections $\sim (m_d-m_u)/\Lambda_\mathrm{QCD}\sim\mathcal O(\%)$. However, this contribution to isospin breaking is actually even further suppressed: thanks to heavy-quark symmetry, the $\bar B\to D^{(*)}$ matrix element exhibits a normalization independent of the spectator quark at the endpoint ($q^2=(p_{\bar B}-p_{D^{(*)}})^2=(m_{\bar B}-m_{D^{(*)}})^2$) in the heavy-quark limit \cite{Isgur:1989vq, Isgur:1990yhj}, implying a further suppression of isospin breaking $\sim 1/m_{c,b}$ at this point. Away from this kinematical configuration, the fact that the form factors can be expanded in the dimensionless quantity $z$, with $|z|\lesssim 6\%$ and the coefficients in this expansion being limited by unitarity (see, for instance, Refs.~\cite{Boyd:1995sq,Boyd:1997kz} and references therein), provides suppression of isospin breaking even away from the endpoint, as discussed for instance in Ref.~\cite{Kobach:2019kfb} for the case of $SU(3)$ symmetry. As a result, the expected symmetry breaking from this source is only at the per mil level.

The second source for isospin breaking are electromagnetic interactions, since the different charges of the spectator quarks, $q_u\neq q_d$, cause differences in interactions with photons. These are generically expected to scale as $\alpha/\pi\sim 0.2\%$, and in this case there is a priori no reason to expect further suppression from heavy-quark symmetry. On the contrary, 
Ref.~\cite{Atwood:1989em} finds an enhancement of this breaking, which becomes maximal at the endpoint of the electron-energy spectrum and at this point yields an additional factor of $\pi^2\sim 10$. Since the precise energy dependence for the full spectrum is not known, we conservatively allow a variation in the isospin relation in Eq.~\eqref{eq::isospin} by this maximal value  $\pi^2(\alpha/\pi)\sim2\%$, independently for $\bar B\to D$ and $\bar B\to D^*$.

\section{Analysis and Results \label{sec:results}}

\subsection{Treatment of common input parameters and correlations \label{sec:treatment-common}}
\newcounter{mpFootnoteValueSaver}

Given the dominance of systematic uncertainties in a majority of measurements, which is expected to continue and grow with the coming even larger datasets, the goal is to include as many of the resulting correlations as possible. Within a given measurement, this task is usually performed within the analysis, but we aim to include also correlations between different measurements, similar to what is being done by HFLAV~\cite{HeavyFlavorAveragingGroupHFLAV:2024ctg}. These correlations largely result from the dependencies detailed in Eq.~\eqref{eq::Nexp} and can correspondingly be incorporated. This allows in principle to take into account the correlations due to $B$ production, secondary branching fractions, and lifetimes (which do not enter Eq.~\eqref{eq::Nexp} explicitly, but appear in cases where the branching fraction of the semileptonic decay is explicitly parametrized, for instance using the CLN parametrization \cite{Caprini:1997mu}). The difficulty is that the available information is sometimes limited, especially in older measurements. When sufficient
information is provided by the experiment, we can account for the correlations and furthermore undo approximations or assumptions that were applied in the measurement (like exact isospin symmetry for the $B$ decay rates or equal production fractions for charged and neutral $B$ mesons), and correct for d'Agostini bias, as explained below. We proceed generally as follows, similarly to our analysis of $B\to \bar D D$ decays \cite{Davies:2023arm}:
We determine an \emph{effective counting rate} for each measurement via
    \begin{align}
        N_\mathrm{eff}(B^{\pm,0}\to X(\to Y))&\equiv \frac{N(B^{\pm,0}\to X(\to Y))}{\epsilon_{B^{\pm,0}\to X(\to Y)}}
        = N_{B^{\pm,0}} BR(B^{\pm,0}\to X)BR(X\to Y)\,,
    \end{align}
where both expressions on the right can be used to obtain the result, again depending on the information provided in the analysis. If the second expression is applied, the input values used in that specific analysis have to be inserted. In the resulting effective counting rate the systematic uncertainty is correspondingly reduced compared to the one of the BR given in the analysis, since the uncertainties of the external quantities are now separately accounted for. 
If several quantities are provided by the same measurement, their correlations have to be taken into account. The important property of 
the effective counting rate
is that it is approximately independent from external inputs.\footnote{There are commonly implicit  dependencies, for example through the experimental efficiencies.} Providing this number in experimental analyses would therefore allow to account for changes in external 
inputs easily. If the experimental result is provided in terms of a ratio with a normalization mode, the corresponding effective quantity is calculated analogously.
The resulting effective counting rates are then fitted with up-to-date values for the external inputs~\cite{ParticleDataGroup:2024cfk,HeavyFlavorAveragingGroupHFLAV:2024ctg}, which implicitly performs the rescaling done by HFLAV. Given the form of the inputs, the correlations following from external inputs are taken into account automatically. Up to this point the approach is general, \emph{i.e.}, applicable to any branching-fraction measurement.
Within this general approach, a couple of subtleties arise specifically for the semileptonic modes in question which we discuss in the following.

One complication arises in measurements that use different $D$ decay modes for a given
$B$ decay when
the relative efficiencies for these different final states are 
not provided. This renders the precise inclusion of updated branching fractions for the $D$ decays difficult or impossible. Our treatment of these cases is described in detail 
in Appendix~\ref{app:multiD}.
  
The treatment of the branching fraction of $D^0\to K^-\pi^+$ deserves special attention, given its small uncertainty and important role both in \BD and \BDs decays. HFLAV has devoted an extended analysis to this \cite{HeavyFlavorAveragingGroupHFLAV:2024ctg}, emphasizing the importance of the consistent treatment of additional photons in these measurements. This aspect should be explicitly considered in future analyses. Below we nevertheless continue to use the branching fraction provided by the PDG~\cite{ParticleDataGroup:2024cfk}, for two reasons:
(i)~The PDG provides the correlations with other secondary $D$ decays, which are taken into account by us where possible.
(ii)~Even if we agree that a consistent treatment of this branching fraction is important, it is not clear that the one chosen by HFLAV corresponds to the one chosen in the \BDDs measurements under consideration. It is therefore not obvious that the HFLAV branching fraction improves our analysis.

In a couple of cases a measurement concerns charged and neutral $B$ decays, but the correlations for their systematic uncertainties are not explicitly provided. If both the individual results and a combined BR are given, it is possible to use this information to estimate an \emph{effective correlation} between the individual results, \emph{i.e.}, the combined result is calculated from the individual ones as a weighted average with an a priori unknown correlation, and the comparison with the given combined result determines the correlation coefficient. 
The reason we call this an \emph{effective} correlation is due to the fact that considering just the BRs amounts to the reduction of several multi-parameter fits to the limited subspace of branching fractions. In general, it is \emph{not} possible to represent the full information of such fits by a simple correlation matrix. For instance, many of the listed measurements obtain the branching ratio as an integral over the differential rate expressed in the CLN parametrization. If the results for the charged and the neutral mode are obtained in separate analyses, the corresponding branching fractions will be implicitly correlated by the fact that both are parametrized in terms of CLN parameters. However, the central values of each result will correspond to different values for those parameters. Performing then a joint fit with a single set of CLN parameters would shift the result for the branching fractions in a way that in general cannot be expressed by a simple correlation between the individual BR results. Since, however, we do not have the necessary information to repeat the fits ourselves, using an effective correlation 
is the best approximation available to us.

Finally, we want to detail the way we correct for the d'Agostini bias \cite{DAgostini:1993arp} where possible. This bias arises in fits to experimental data from uncertainties affecting their normalization. In particular, it affects the total branching ratio and correspondingly \Vcb when extracted from a fit to binned differential distributions, as first pointed out in Ref.~\cite{Jung:2018lfu}. Importantly, the effect increases with the number of bins \cite{DAgostini:1993arp}. We correct for this following Refs.~\cite{Bordone:2019vic,Bordone:2019guc} by simply considering the total rate as the sum over all bins of a given distribution (not involving a fit) and the corresponding normalized differential distribution, in which global normalization factors cancel. The advantage compared to the alternative treatment of the d'Agostini bias used for instance in Ref.~\cite{Gambino:2019sif} is that we do not need to separate statistical and systematic covariance matrices, which is information that is not always available. Since the correlation between the normalized bins and the total branching fraction is significantly reduced by this procedure, it also allows to use only the rate information in the following, while the differential information is only used on top of that in the following \Vcb analysis~\cite{Gambino:inpreparation}. 
 \begin{figure}[t]
        \centering{
        \includegraphics[width=9cm]{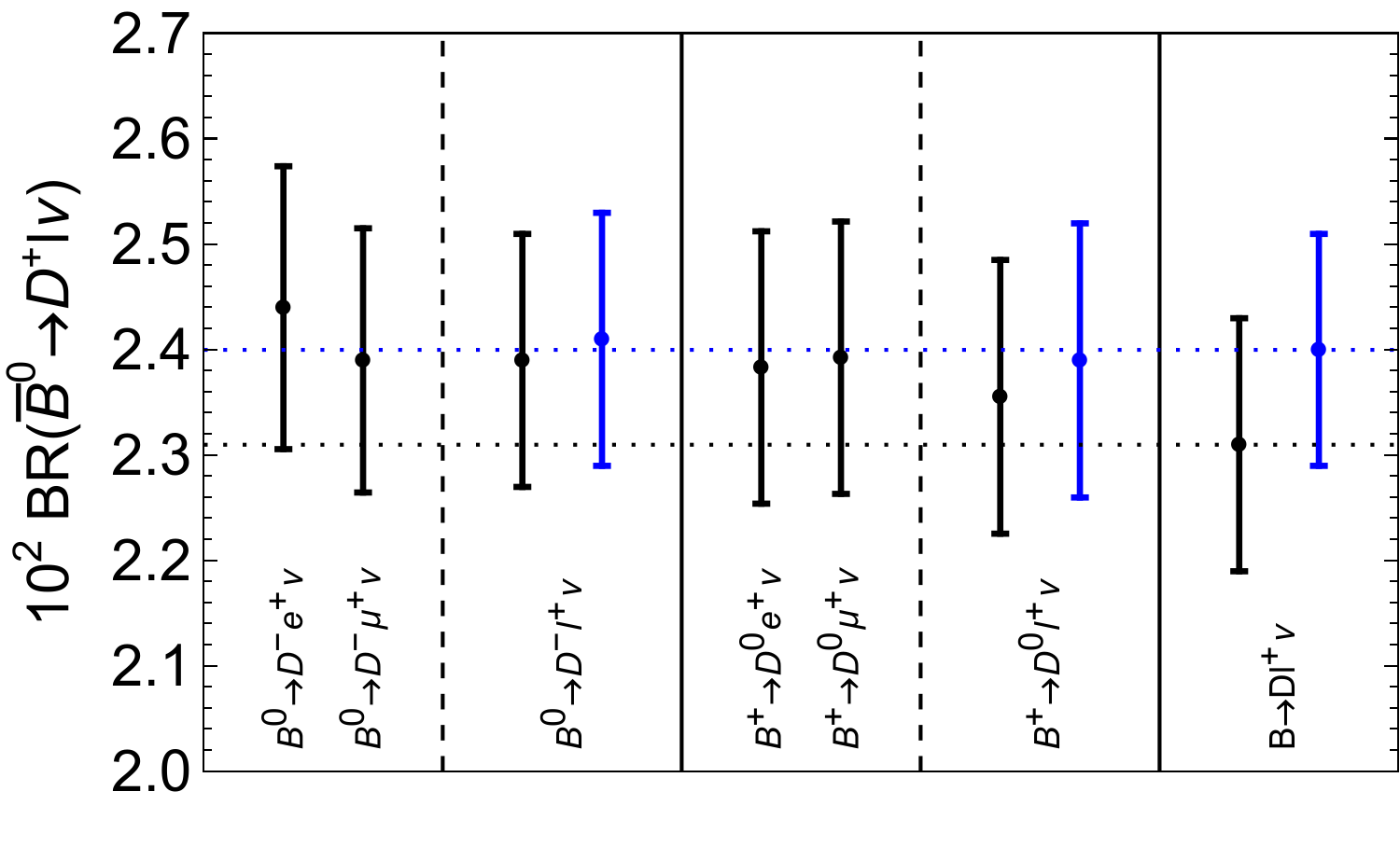}}
        \caption{%
        \setcounter{mpFootnoteValueSaver}{2}%
        $B\to D$ branching fractions obtained from the Belle'15 measurement~\cite{Belle:2015pkj}, multiplied by the lifetime ratio in case of the charged $B$ decays for easier comparison. In black the results from the paper, reproduced by us,\protect\footnotemark[\value{mpFootnoteValueSaver}]%
        {}
        which show that averaging the different modes leads to smaller and smaller values. The fits to the normalized data yield the same results for the individual decay modes, while the averages are consistently higher, as expected when correcting for d'Agostini bias. Note that here the external inputs from Ref.~\cite{Belle:2015pkj} are used consistently in order to ensure comparability, \emph{i.e.}, the values are not rescaled at this point (in contrast to the values given in Table~\ref{tab:B2Dresults}).          
        \label{fig::BtoDBelledAgostini}
        }
    \end{figure} 

To illustrate the effect beyond the observations made already in Refs.~\cite{Jung:2018lfu,Gambino:2019sif}, for instance, we show in Fig.~\ref{fig::BtoDBelledAgostini} the branching ratio for \BD obtained from the measurement in Ref.~\cite{Belle:2015pkj}, averaging over different numbers of bins: the lepton- and isospin-specific measurements are in both cases just given by the sum over 10 bins provided for each channel in that reference and hence do not differ between Ref.~\cite{Belle:2015pkj} and our analysis. However, when fits are performed in order to obtain the lepton-flavour average or the combined lepton-flavour and isospin average, we observe how the averaged branching fractions obtained in Ref.~\cite{Belle:2015pkj} displayed in black become smaller than the individual branching fractions entering the averages, with the effect becoming largest in the final average involving all 40 bins. Performing the same averages, but using the normalized differential distributions and total rates instead, we obtain the branching fractions displayed in blue, showing the expected behaviour instead. For further details of this measurement, see the discussion in Appendix~\ref{sec:data-Belle-15}. 

Note that when analyses obtain the branching fraction from a fit to the CLN parametrization, the results thus suffer usually from \emph{two} shortcomings: first, the parametrization itself is insufficient to describe the form factors properly at the present level of precision, since it makes approximations that are not warranted anymore \cite{Bigi:2016mdz,Bernlochner:2017jka, Bigi:2017njr, Grinstein:2017nlq, Jaiswal:2017rve, Bordone:2019vic}. Even when the data are well described by this model, the related uncertainties are underestimated by varying the fit parameters, and an additional uncertainty for this is not assigned. Second,  these fits are performed without accounting for d'Agostini bias. Therefore the results from such analyses can be expected to be biased towards too small central values \emph{and} uncertainties.

\footnotetext[\value{mpFootnoteValueSaver}]{We reproduce these values assuming lepton-flavour universality (LFU) for each of the decay modes separately in the second and fourth panels. Using LFU for both $\bar B^0\to D^+\ell\bar\nu$ and $B^-\to D^0\ell\bar\nu$ results in even lower values.}
\addtocounter{footnote}{1}

To judge the overall fit quality of the averages and specifically the compatibility among the various \BDDs measurements, we provide both the overall $\chi^2$/dof and $\Delta\chi^2/\Delta$dof, obtained as the difference to a baseline fit including external inputs, only.  
$\Delta\chi^2/\Delta$dof is expected to provide a better measure for the consistency of the branching-ratio data, only. This is useful to separate potential peculiarities in the external inputs, which are secondary to our analysis, from the main parameters of interest.
For instance, our external input for the production fractions at the $B$ factories from Ref.~\cite{Bernlochner:2023bad} provides effectively 3 observables for 2 free parameters $R^{\pm0}$ and 
$f_{\slashed{B}}$ that contribute very little to the overall $\chi^2$.\footnote{We parametrize the dependence on the production fractions in terms of $R^{\pm0}$ and $f_{\slashed{B}}$ in order to minimize correlations.}  Given this situation, the total $\chi^2$$/\text{dof}$ tends to be on the low side for any fit including this input.

Apart from the slightly different treatment of  external input parameters and correlations detailed above, the main differences to the HFLAV average in Ref.~\cite{HeavyFlavorAveragingGroupHFLAV:2024ctg} are the following, see Appendix~\ref{app:BRdetails} for more details: 
\begin{enumerate}
    \item Corrected treatment of the branching fraction information from the CLEO measurements~\cite{CLEO:1998qvx,CLEO:2002fch,CLEO:2002vsd}. 
    \item Consistent treatment of the inclusive branching ratio for \BX used in Refs.~\cite{BaBar:2007ddh,BaBar:2009zxk}. 
    \item Accounting for d'Agostini bias in Refs.~\cite{Belle:2015pkj} and \cite{Belle:2018ezy}, as discussed above, see Fig.~\ref{fig::BtoDBelledAgostini}.  
    \item Inclusion of the Belle-II measurement \cite{Belle-II:2025rna}, which supersedes the preliminary results given in the proceeding article~\cite{Belle-II:2022ffa}.
\end{enumerate}

\subsection{Experimental information on \BD}

\begin{figure}[t]
    \centering{\includegraphics[width=9cm]{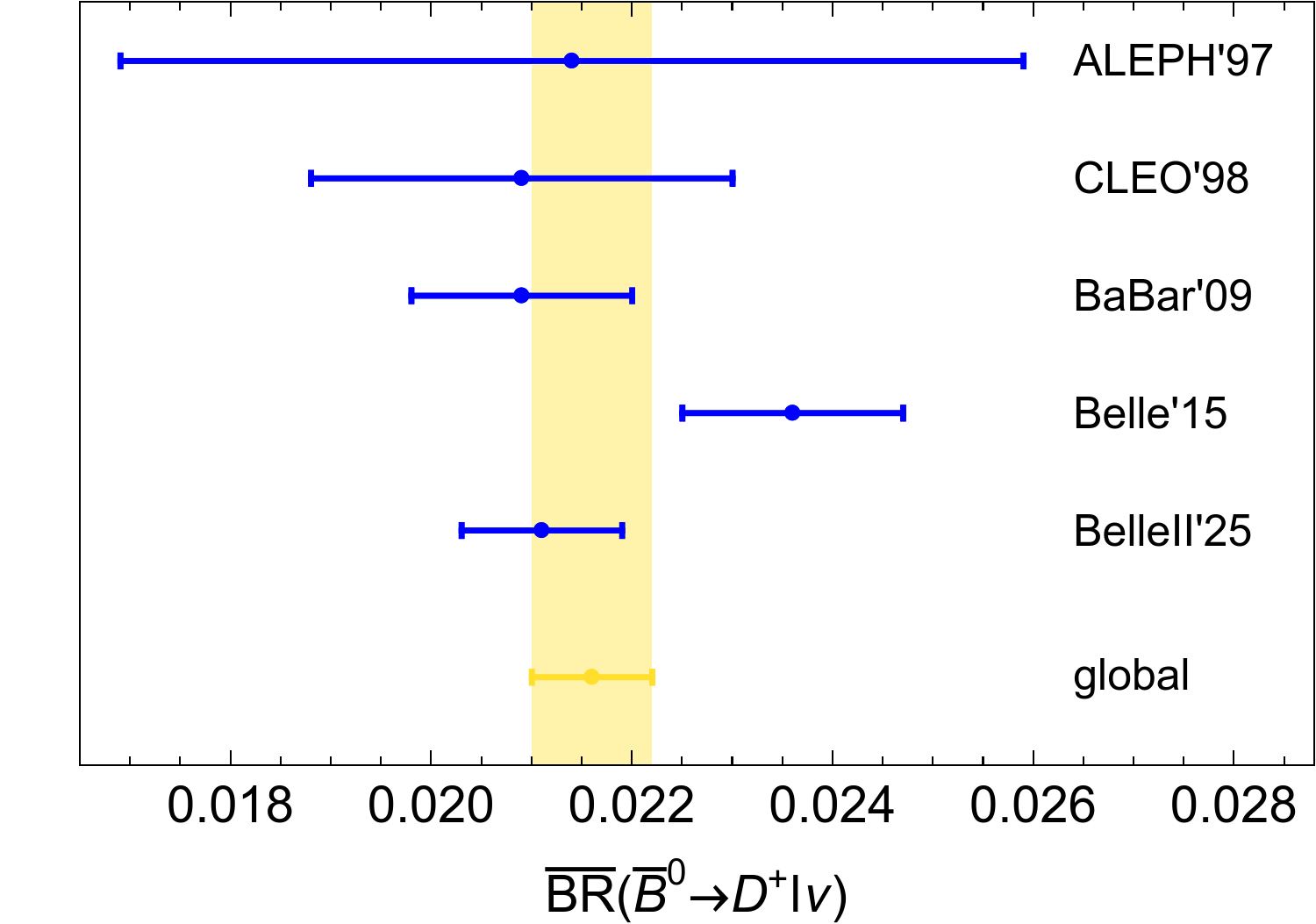}}
    \caption{Results for the isospin-averaged branching fraction expressed as $BR(\bar B^0\to D^+\ell\bar\nu)$ from the individual measurements as well as the result from the global \BD fit, see Table~\ref{tab:B2Dresults}. 
    \label{fig::BRBD}}
\end{figure}

\begin{figure*}[t]
\centering{
\includegraphics[width=0.55\linewidth]{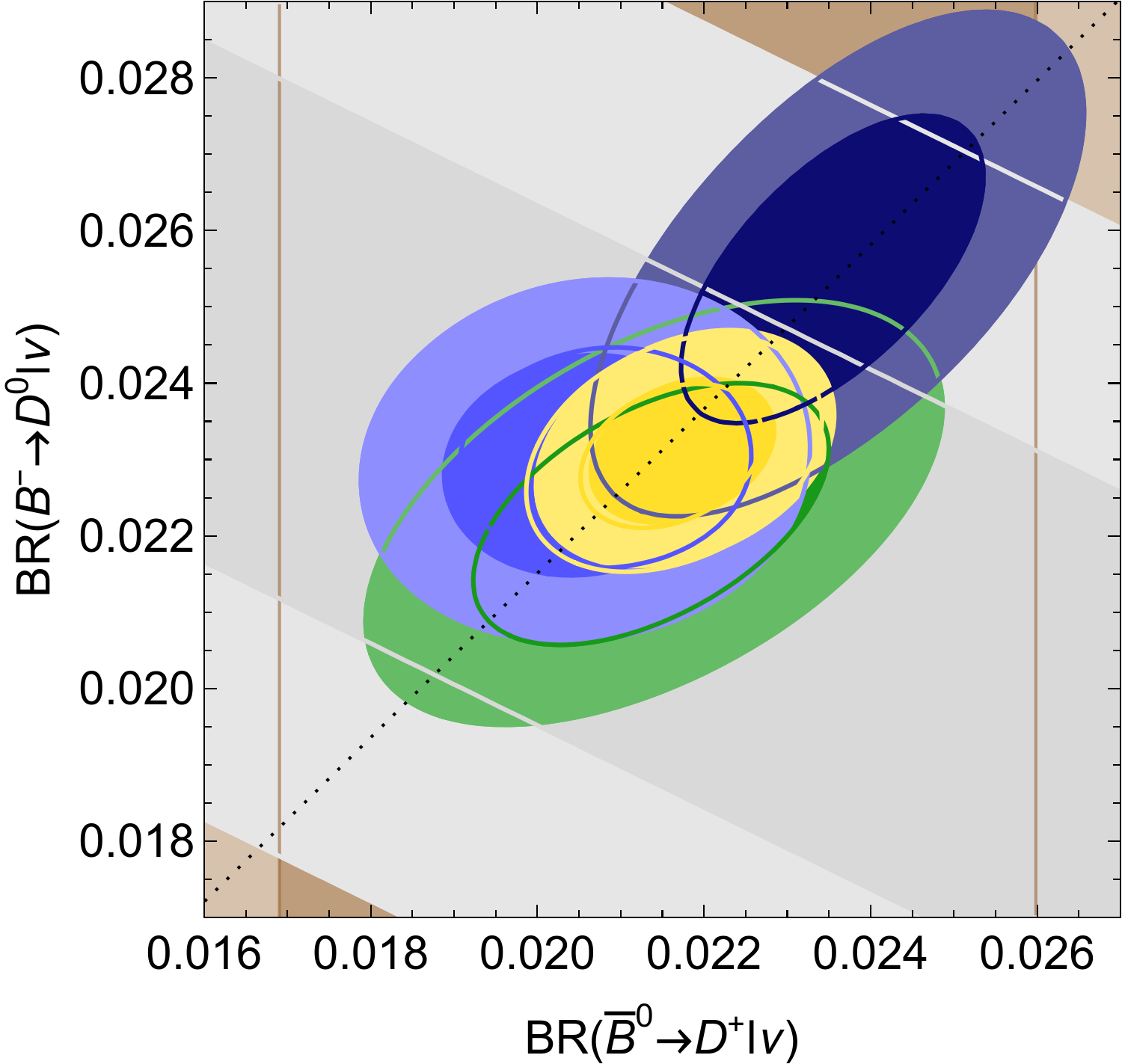}
}
\caption{ \label{fig::BDglobal} Global fit (yellow) to the available \BD measurements. Shown are the constraints at 68$\%$ and 95$\%$ CL from ALEPH'97 (brown), CLEO'98 (grey), BaBar'09 (green), Belle'15 (blue), and Belle~II'25 (light blue). The dotted line corresponds to the isospin limit for the two rates.}
\end{figure*}

We start by analyzing the data from \BD{} alone. 
We perform fits treating the branching fractions of neutral and charged $B$ decays independently, before we combine them assuming approximate isospin symmetry, assigning an uncertainty to this assumption as described in Sec.~\ref{sec:isospin-B-decays}. 
The results of these fits are collected in Table~\ref{tab:B2Dresults}, and illustrated in Figs.~\ref{fig::BRBD} and~\ref{fig::BDglobal}. There we also present the rescaled values for the individual branching fractions as inferred from the effective observables entering our analysis. A recent BaBar analysis \cite{BaBar:2023kug} does not provide independent branching-fraction information and is therefore not included.

All measurements are compatible with isospin symmetry within the given uncertainties, and consequently so are their averages. The overall consistency between the measurements is excellent, as indicated for each fit by both $\chi^2/$dof and $\Delta\chi^2/\Delta$dof and illustrated in Figs.~\ref{fig::BRBD} and~\ref{fig::BDglobal}. The Belle'15 measurement is slightly on the high side, which is not a surprise with respect to the older measurements, where the d'Agostini bias could not be corrected for, while the difference with the Belle-II measurement cannot be explained this way, since that result already accounts for this effect. We obtain a sizable correlation between the charged and neutral decays.

\begin{table*}[t]
    {\centering{
\resizebox{\textwidth}{!}{
    \begin{tabular}{l c c c l}\hline\hline
        Measurement & $10^2BR(\bar B^0\to D^+\ell\bar\nu)$  & $10^2BR(B^-\to D^0\ell\bar\nu)$   & $10^2BR(\bar B\to D\ell\bar \nu)$ & Comments\\\hline\hline
        ALEPH'97~\cite{ALEPH:1996dlo}       & 2.14(45)  & ---       & 2.14(45)   & Uses $BR(\bar B\to D^*\ell\bar\nu)$.\\
        CLEO'98~\cite{CLEO:1998qvx}       & ---       & ---       & 2.09(21)  & No separate measurement\\
        BaBar'09~\cite{BaBar:2009zxk}       & 2.13(14)   & 2.23(11)  & 2.09(11)   & Corrected \BX interpretation\\
        Belle'15~\cite{Belle:2015pkj}      & 2.35(12)    & 2.55(13)    & 2.36(11)   & Re-fit to account for d'Agostini bias\\
        Belle-II'25~\cite{Belle-II:2025rna}   &  2.08(12)  & 2.30(9) &  2.11(8) & \\\hline
        Average         & 2.17(7)     & 2.31(6)   & 2.16(6)   & Correlation($\bar B^0,B^-$) $29.0\%$  \\\hline
        $\chi^2/$dof    & \multicolumn{2}{c}{5.0/7} & 5.0/8  &\\
        $\Delta\chi^2/\Delta$dof    & \multicolumn{2}{c}{4.9/6}  & 4.9/7  &\\\hline
       \hline
    \end{tabular}
}}
    \caption{Results for the rescaled experimental \BD 
    branching ratios and our averages.\label{tab:B2Dresults}  
    The branching fractions in the isospin limit are given as the common rate  multiplied by the neutral-meson lifetime.}
    }
\end{table*}

The differences of our results with the HFLAV averages~\cite{HeavyFlavorAveragingGroupHFLAV:2024ctg} amount to shifts with different signs, but with the more sizable shifts tending to increase the obtained branching fractions. In particular, the single largest correction is the one from correcting the results of Ref.~\cite{Belle:2015pkj} for d'Agostini bias, illustrated in Fig.~\ref{fig::BtoDBelledAgostini}, see also Table~\ref{tab:B2Dresults}.
The resulting averages are about $0.7\sigma$ higher for \BRBzeroD and slightly more than $1.6\sigma$ higher for \BRBminusD; the isospin average increases by $\sim 0.8\sigma$.

\subsection{Experimental information on \BDs}

\begin{figure}[t]
{\centering{\begin{minipage}[t]{9.cm}\vspace*{0mm}
\includegraphics[width=9cm]{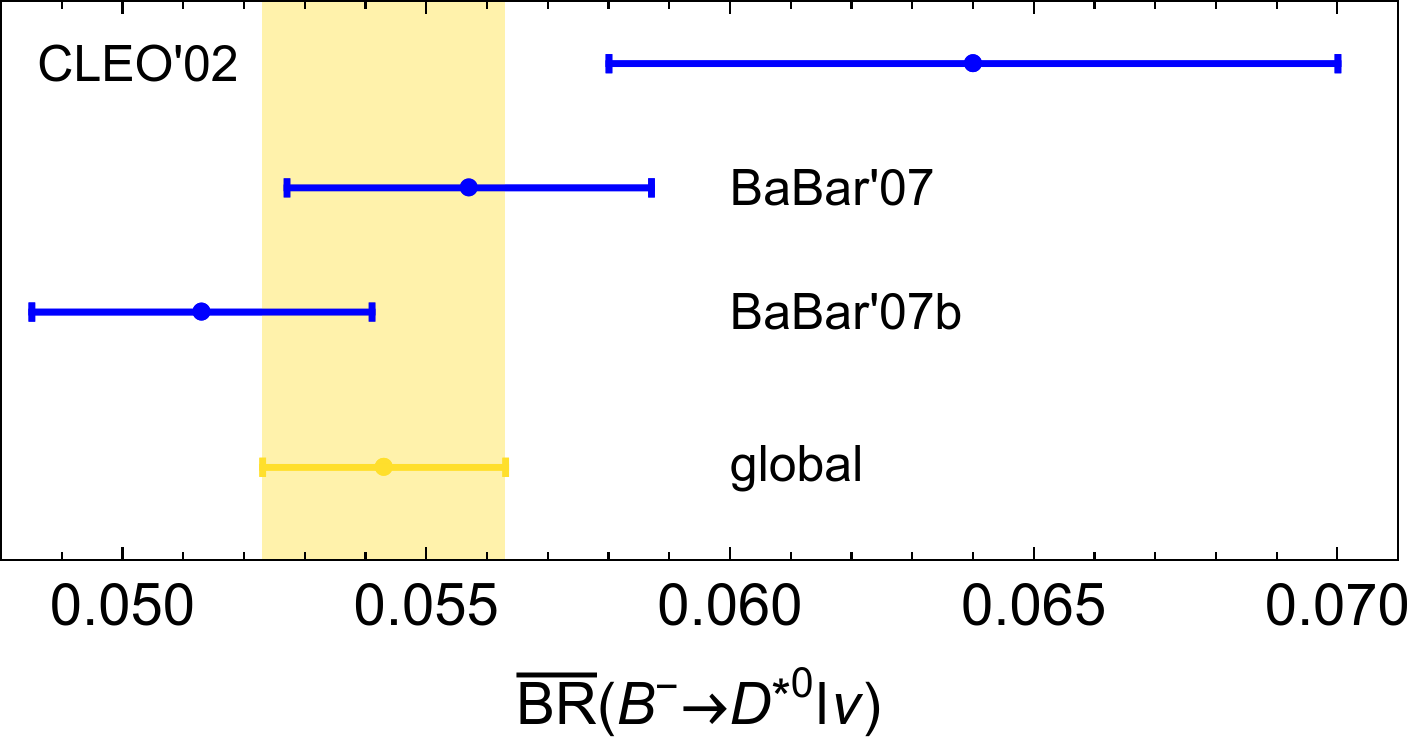} 
\end{minipage}\hfill
\begin{minipage}[t]{5.5cm}\vspace*{0mm}
\begin{tabular}{ll}\hline\hline Measurement & $BR/\%$\\\hline
    CLEO'02   & $6.4(6)$   \\
    BaBar'07a  & $5.57(30)$ \\
    BaBar'07b & $5.13(28)$ \\\hline
    Global $B^-\rightarrow D^{*0}\ell\bar\nu$    & $5.43(20)$   \\\hline
    $\chi^2/\mathrm{dof}$ &  4.1/3   \\
     $\Delta\chi^2/\Delta\mathrm{dof}$  & $4.0/2$ \\\hline\hline
    \end{tabular}
\end{minipage}
}}
    \caption{Results for $BR(B^-\to D^{*0}\ell\bar\nu)$, only, from the individual measurements as well as the result from their combination.
    \label{fig::BRBcDstar}}  
\end{figure}

\begin{figure}[t]
{\centering{
\begin{minipage}[t]{9.cm}\vspace*{0mm}
\includegraphics[width=9cm]{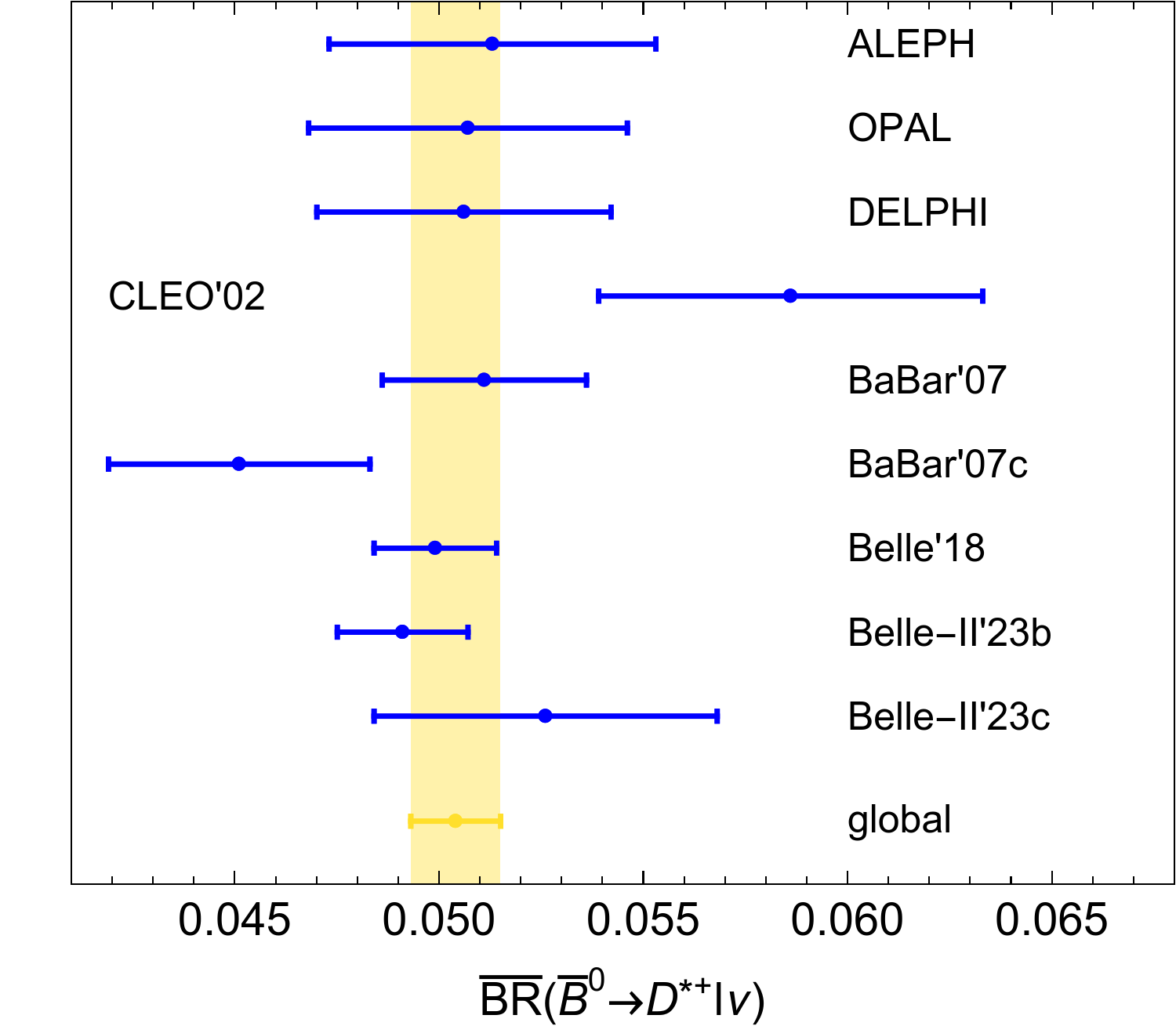} 
\end{minipage}\hfill%
\begin{minipage}[t]{5.7cm}\vspace*{0mm}
\begin{tabular}{ll}\hline\hline Measurement & $BR/\%$\\\hline
    ALEPH'97    & $5.13(40)$\\
    OPAL'00     & $5.07(39)$\\
    DELPHI'04   & $5.06(36)$\\
    CLEO'02   & $5.86(47)$  \\
    BaBar'07a  & $5.11(25)$ \\
    BaBar'07c & $4.51(32)$ \\
    Belle'18    & $4.99(15)$ \\
    Belle-II'23b& $4.91(19)$\\
    Belle-II'23c& $5.26(42)$ \\\hline
    Global  $\bar B^0\to D^{*+}\ell\bar\nu$   & $5.02(11)$ \\\hline
    $\chi^2/\mathrm{dof}$   &  9.6/11 \\
    $\Delta\chi^2/\Delta\mathrm{dof}$   &  9.5/10 \\\hline\hline
    \end{tabular}
\end{minipage}
}}
    \caption{Results for $BR(\bar B^0\to D^{*+}\ell\bar\nu)$, only, from the individual measurements as well as the result from their combination.
    \label{fig::BRB0Dstar}}
\end{figure}

\begin{figure}[t]
    \centering{
    \includegraphics[width=10cm]{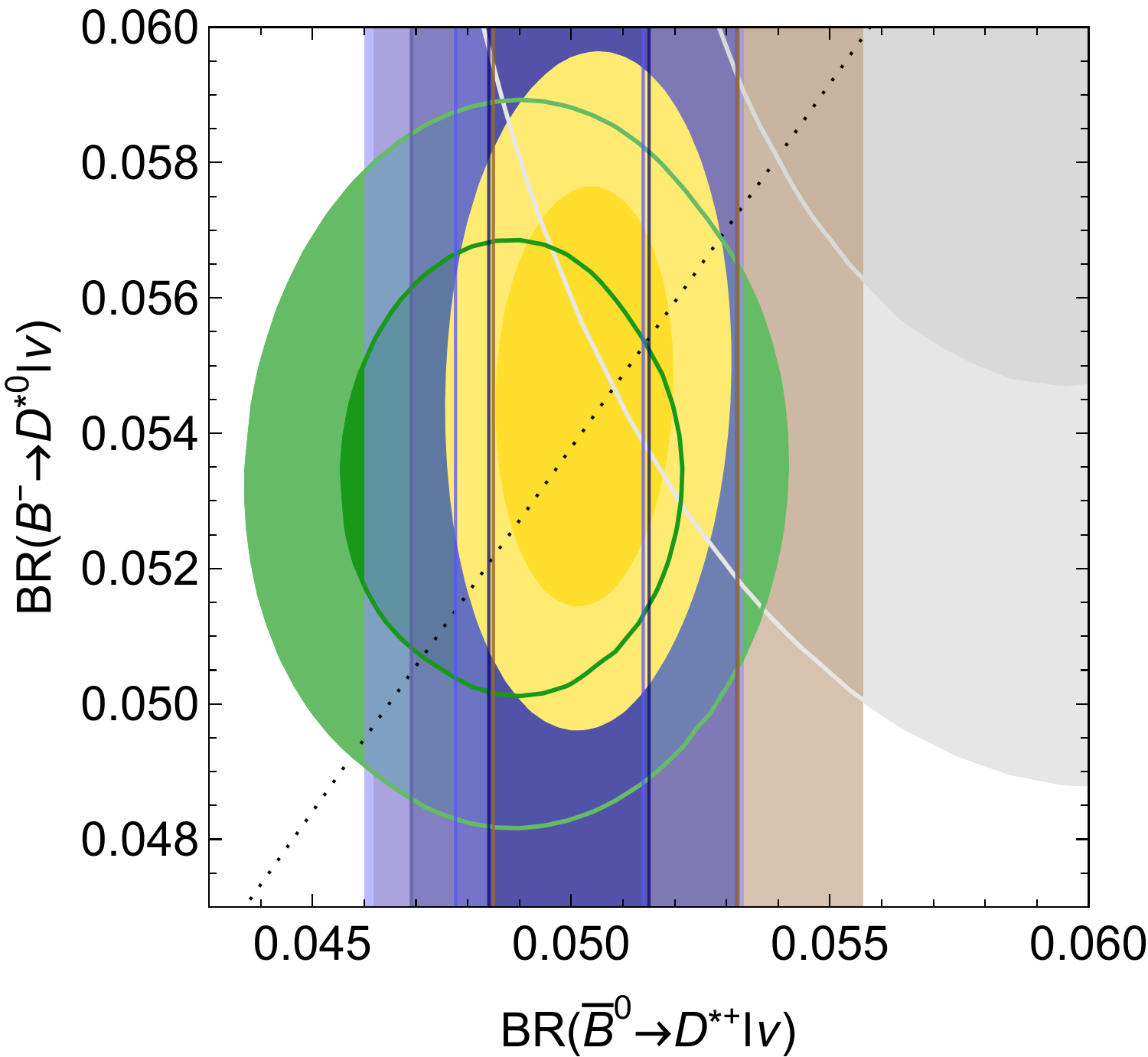}}
    \caption{
    Global fit (yellow) to the available \BDs measurements. Shown are the combined constraints for each experiment at 68$\%$ and 95$\%$~CL from LEP (brown), CLEO (grey), BaBar (green), Belle (dark blue), and Belle~II (light blue). The dotted line corresponds to the isospin limit for the two rates.
    \label{fig::BRBDstar}}
\end{figure}

For \BDs{} we proceed analogously to \BD{}. We show our results for the rescaled branching ratios and the corresponding averages in Table~\ref{tab:B2Dstarresults} and Figs.~\ref{fig::BRBcDstar}--\ref{fig::BRBDstar}. One more Belle-II analysis, Belle-II~23a \cite{Belle-II:2023svm}, and two more Belle analyses, Belle~23a~\cite{Belle:2023bwv} and Belle~23b~\cite{Belle:2023xgj}, are available, but do not provide the branching fractions and are hence not included in the following. 
As can be seen from Table~\ref{tab:B2Dstarresults} and Fig.~\ref{fig::BRBDstar}, the overall fit quality is again very good. The separate results for charged and neutral $B$ decays are shown in Figs.~\ref{fig::BRBcDstar} and \ref{fig::BRB0Dstar}, respectively, displaying no significant tensions, neither among the $B^-$ nor the $\bar B^0$ measurements.
We observe again excellent compatibility with the isospin-symmetry limit. The net correlation, resulting from combining large positive and negative correlations, is smaller than for \BD. 

Compared to HFLAV, our averages exhibit significantly higher confidence levels, due to the differences described above. Our average for $BR(\bar B^0\to D^{*+}\ell\bar \nu)$ is about $1\sigma$ higher, mostly due our re-fitted result of the measurement in Ref.~\cite{Belle:2018ezy} 
analogous to the one of Ref.~\cite{Belle:2015pkj}, see Sec.~\ref{sec:treatment-common}. On the other hand, the average
for $BR(B^-\to D^{*0}\ell\bar \nu)$ is slightly lower, due to our different interpretation of the CLEO result in Refs.~\cite{CLEO:2002fch,CLEO:2002vsd}. 
Assuming isospin symmetry in the same way as in $B\to D$, the combined average is $1.2\sigma$ higher in comparison.

\begin{table*}[t]
    \centering
\resizebox{\textwidth}{!}{
    \begin{tabular}{l c c c l}\hline\hline
        Measurement & $\bar B^{0}\to D^{*+}\ell\bar\nu$  & $B^-\to D^{*0}\ell\bar\nu$   & $\bar B\to D^*\ell\bar \nu$ & Comments\\\hline\hline
        ALEPH'97~\cite{ALEPH:1996dlo}       & 5.13(40)    & ---       & 5.13(40)    & Correlated with $\bar B\rightarrow Dl\bar\nu$ \\
        OPAL'00~\cite{OPAL:2000hcv}       & 5.07(39)   & ---       & 5.07(39)  & Two correlated methods for $D$ reconstruction \\
        DELPHI'01/04 \cite{DELPHI:2001def, DELPHI:2004hkn}   & 5.06(36)   & ---       & 5.06(36)  &  Two correlated methods for $D$ reconstruction\\
        CLEO'02~\cite{CLEO:2002fch}       & 5.86(47)     & 6.4(6) & $5.90(35)$  & Separate treatment of $B^{0,-}$ decays. \\
        BaBar'07a~\cite{BaBar:2007ddh}      & 5.11(25) & 5.57(30)  & $5.14(20$)  & Corrected \BX interpretation\\
        BaBar'07b~\cite{BaBar:2007nwi}     & --- & 5.13(28) & $4.77(28)$ & CLN parameters rescaled  \\
        BaBar'07c~\cite{BaBar:2007cke}     & $4.51(32)$ & --- & $4.51(32)$ & Allows for check of $D$ rescaling\\
        Belle'18~\cite{Belle:2018ezy}      & 4.99(15)  & ---   & 4.99(15)  & Re-fit to account for d'Agostini bias\\
        Belle-II'23b~\cite{Belle-II:2023okj}   & $4.91(19)$  & ---  & $4.91(19)$  &  \\
        Belle-II'23c~\cite{Belle-II:2023jtw}     & $5.26(42)$   &--- & 5.26(42)  & Preliminary \\\hline
        Average         & 5.02(11)   & 5.45(21)   & 5.03(11)  & Correlation($\bar B^0,B^-$) $11.5\%$ \\\hline
        $\chi^2$/dof    & \multicolumn{2}{c}{14.7/13}      & 14.7/14 & \\
       $\Delta\chi^2$/$\Delta$dof    & \multicolumn{2}{c}{14.6/12}      & $14.6/13$ &               \\\hline\hline
    \end{tabular}
}
    \caption{Results for the rescaled experimental \BDs{} branching fractions  and our averages in percent. The third column  shows the isospin limit result for $\bar B^0$ decays, given again as the common rate  multiplied by the neutral-meson lifetime.}    
    \label{tab:B2Dstarresults}
\end{table*}

\subsection{Determination of production fractions and global fit}

A key feature of our analysis is its sensitivity to the $B$-production parameters $R^{\pm0}$ and $f_{\slashed{B}}$. Given the difficulty to extract independent information on the production fractions mentioned above, it is important to understand where the sensitivity comes from in this case. 
There are in fact three ways in which we are sensitive to the production fractions:  
\begin{enumerate}
    \item The first is free from explicit isospin-symmetry assumptions, and uses the fact that we obtain an absolute branching fraction from $Z$ decays, using the $B$ production fraction for neutral $B$ mesons measured at LEP.\footnote{This measurement assumes isospin symmetry for neutral and charged $B$-meson production at the $Z$ pole, however. This assumption can be reasonably expected to hold since the $Z$ boson decay is nowhere close to any thresholds relevant to $B$ production.} Given the large uncertainties of the corresponding measurements, this gives a rather weak bound.
    \item The measurements of ratios with inclusive semileptonic decays~\cite{BaBar:2007ddh,BaBar:2009zxk} are independent of the corresponding production fraction; to obtain the branching fraction an assumption has to be made regarding the normalization modes, see Appendix~\ref{app:BRdetails} for details. However, since isospin breaking is suppressed in inclusive decays \cite{Gronau:2006ei,Bernlochner:2023bad}, this is a reasonable approximation. 
    The combined information from this and the previous point allows to obtain information on $B$ production without using external information on $f_{\pm,00}$ and without assuming isospin symmetry for the $B\to D^{(*)}\ell\bar\nu$ transitions:
    \begin{equation}
        R^{\pm0} = \frac{f_\pm}{f_{00}} = 1.09(5)\,,\label{eq:Rpm0-no-external-isospin}
    \end{equation}
    which is $1.9\sigma$ from unity. 
    \item The third possibility is to assume approximate isospin symmetry for \BDDs, as discussed in detail in Sec.~\ref{sec:isospin-B-decays}. 
    Including this approximation with the added uncertainty discussed above, we obtain instead
    \begin{equation}
        R^{\pm0} = 1.072(35)\,, \label{eq:Rpm0-no-external-isospin-limit}
    \end{equation}
    with a smaller central value, but also a significantly reduced uncertainty, resulting in a significance of $2.2\sigma$. 
    This determination 
    is competitive and actually slightly more precise than the other determinations from other channels entering the present world average \cite{HeavyFlavorAveragingGroupHFLAV:2024ctg,Bernlochner:2023bad}.
\end{enumerate}

\begin{table*}[t]
    {\centering{
    \begin{tabular}{lcc}\hline\hline
       Mode  & no isospin & isospin\\\hline\hline
       $\chi^2/\mathrm{dof}$            & 19.2/16   &  19.6/18 \\
    \hline      
       $\bar B^0\to D^{*+}\ell\bar\nu$  & 5.08(16)   & 5.06(14) \\ 
       $B^-\to D^{*0}\ell\bar\nu$       &  5.42(24) &  \\
       $\bar B^0\to D^{+}\ell\bar\nu$   & 2.20(9)  & 2.17(8) \\
       $B^-\to D^{0}\ell\bar\nu$        & 2.29(9)   &   \\
        $R^{\pm0} $                    &  1.09(5)   & 1.072(35)\\
        $f_{\slashed{B}}$                 &  0.006(30)  &  0.009(30)  \\\hline\hline
    \end{tabular}
\parbox{6cm}{\begin{align*}
\mathrm{corr}_{\mathrm{no-iso}}&=   
\left(
\begin{array}{cccccc}
 1 & 0.185 & 0.504 & 0.236 & 0.362 & 0.569 \\
 0.185 & 1 & 0.119 & 0.488 & -0.406 & 0.513 \\
 0.504 & 0.119 & 1 & 0.331 & 0.295 & 0.449 \\
 0.236 & 0.488 & 0.331 & 1 & -0.539 & 0.671 \\
 0.362 & -0.406 & 0.295 & -0.539 & 1 & -0.277 \\
 0.569 & 0.513 & 0.449 & 0.671 & -0.277 & 1 \\
\end{array}
\right) \\
\mathrm{corr}_{\mathrm{iso}}&=\left(
\begin{array}{cccc}
1 & 0.507 & 0.163 & 0.693 \\
 0.507 & 1 & -0.082 & 0.639 \\
 0.163 & -0.082 & 1 & -0.233 \\
 0.693 & 0.639 & -0.233 & 1 \\
\end{array}
\right)
\end{align*}
}
    \caption{
    Results from the global fits to \BDDs data without external inputs on $R^{\pm0}$ and $f_{\slashed{B}}$. All branching fractions are given in percent.
    }
 \label{tab::BRresults-no-ext}}}
\end{table*}

In Table~\ref{tab::BRresults-no-ext} we present the fit results from the global fits corresponding to the results in Eqs.~\eqref{eq:Rpm0-no-external-isospin}, \eqref{eq:Rpm0-no-external-isospin-limit}: we employ the full information from \BDDs with updated external inputs \cite{ParticleDataGroup:2024cfk,HeavyFlavorAveragingGroupHFLAV:2024ctg},   
again with and without the additional assumption of isospin symmetry.
In this fit, we do not include
the external input on the production fractions, for two reasons: first, it allows to assess the impact of the new modes by themselves, and second, this result can be combined with any set of other measurements for \Rpmz and/or $f_{\not{B}}$, to obtain updated results on the \BDDs branching fractions.
In the fit assuming isospin symmetry, we assign an uncertainty of $\alpha\pi$ to that assumption, as discussed in Sec.~\ref{sec:isospin-B-decays}. 

The fit quality is again high in both cases, indicating consistency with the isospin limit
\begin{align}
r_I(D^{(*)}) &\equiv \frac{\Gamma(B^-\to D^{(*)0}\ell\bar \nu)}{\Gamma(\bar B^0\to D^{(*)+}\ell\bar \nu)} = 1\,.
\end{align} 
Quantitatively, we obtain for the ratios
\begin{align}
r_I(D) &= 0.969(46)\,, \qquad
r_I(D^*) = 0.992(49)
\end{align}
without imposing isospin symmetry.
The results for the branching ratios and production-fraction parameters show a remarkably high compatibility with the previous fits, due to the similar values for the $B$-production parameters preferred by \BDDs alone and the external input. There is a limited sensitivity to $f_{\not{B}}$, resulting in about twice the uncertainty of the external input for this quantity and fully compatible with it.
The uncertainties of the branching fractions increase compared to fits with the external inputs, between $20-50\%$. The correlations are rather large, both between the different branching fractions and between them and the production parameters, since the corresponding uncertainties are larger in these fits.

\begin{table*}[]
    \centering{
    \begin{tabular}{lcccc}\hline\hline
       Mode  & no isospin & isospin & HFLAV~(no isospin)  & HFLAV~(isospin) \\\hline\hline
       $\chi^2/\mathrm{dof}$            & 19.8/19   & 19.9/21 & & \\
    $\Delta\chi^2/\Delta\mathrm{dof}$   & 19.7/18   & 19.8/20 & & \\\hline
       $\bar B^0\to D^{*+}\ell\bar\nu$  & 5.02(11)   & 5.02(10) & 4.90(12) & 4.90(11) \\ 
       $B^-\to D^{*0}\ell\bar\nu$       & 5.44(20)  & & 5.53(22) & \\
       $\bar B^0\to D^{+}\ell\bar\nu$   & 2.17(7)  & 2.16(6) & 2.12(6) & 2.11(5) \\
       $B^-\to D^{0}\ell\bar\nu$        & 2.31(6) &  & 2.21(6) &  \\
       $R^{\pm0} $                     & 1.063(21) & 1.062(19) & \multicolumn{2}{c}{$1.052\pm 0.031$}   \\
       $f_{\slashed{B}}$                & $0.00264^{+0.012}_{-0.00021}$  &  $0.00264^{+0.012}_{-0.00021}$  &   \multicolumn{2}{c}{$0.00264^{+0.0125}_{-0.0002}$}      \\\hline\hline
    \end{tabular}
\parbox{\textwidth}{
\begin{align*}
\mathrm{corr}_{\mathrm{no-iso}}&= \left(
\begin{array}{ccccc}
 1 & 0.023 & 0.110 & 0.034 & 0.358  \\
 0.023 & 1 & -0.023 & 0.116 & -0.149 \\
 0.110 & -0.023 & 1 & 0.297 & 0.242 \\
 0.034 & 0.116 & 0.297 & 1 & -0.243 \\
 0.358 & -0.149 & 0.242 & -0.243 & 1 
\end{array}
\right)\hfill
\mathrm{corr}_{\mathrm{iso}} = \left(
\begin{array}{ccc}
1 & 0.096 & 0.289 \\
 0.096 & 1 & 0.053 \\
 0.289 & 0.053 & 1  
 \end{array}
\right)
\end{align*}%
}%
\caption{
Results from the global fits to \BDDs data including external inputs on $R^{\pm0}$ and $f_{\slashed{B}}$. The asymmetric uncertainties for $f_{\slashed{B}}$ prevent us from providing its correlation with the other fit parameters, which we checked to be very small, however. For comparison, we also show the corresponding results from HFLAV~\cite{HeavyFlavorAveragingGroupHFLAV:2024ctg}, showing deviations of up to $1.6\sigma$. We do not show the $\chi^2/\mathrm{dof}$ in this case, as HFLAV performs separate fits for $B\rightarrow Dl\nu$ and $B\rightarrow D^*l\nu$ decays, such that a direct comparison with our $\chi^2/\mathrm{dof}$ values is not meaningful. Note that HFLAV does not include the information of \BDDs to determine \Rpmz, $f_{\not{B}}$, hence the given values simply correspond to the external inputs in that case.
}
    \label{tab::BRresults}
    }
\end{table*}

\begin{table}[t]
    {\centering{
    \begin{tabular}{l c c c}\hline\hline
         Scenario       & $R^{\pm0}$ ($\bar B\rightarrow D$)                            
                        & $R^{\pm0}$ ($\bar B\rightarrow D^*$) 
                        & $R^{\pm0}$ (\BDDs )       \\\hline

         No iso, no ext & $1.10(7)$ 
                        & $1.04(8)$ 
                        & $1.09(5)$ \\

         Iso, no ext    & 1.069(44)
                        & $1.05(6)$  
                        &  1.072(35) \\

         No iso, ext    & 1.061(22)                            
                        & $1.058(23)$
                        & 1.063(21) \\

         Iso, ext       & 1.059(21)                         
                        & $1.060(22)$  
                        & $\mathbf{1.062(19)}$ \\
         \hline\hline
    \end{tabular}
    \caption{Values for $R^{\pm0}$ for the individual channels and their combination, obtained in setups with or without external inputs on the production fractions and the assumption of approximate isospin symmetry.  
    The external input corresponds to $R^{\pm0}_\mathrm{ext} = 1.056(23)$~\cite{Bernlochner:2023bad}.
    In this table we impose $f_{\slashed{B}}\geq 0$.
    \label{tab::Rresults}} 
    }}
\end{table}

Adding again the external input on the  production fractions,
we obtain our main results, presented in Table~\ref{tab::BRresults}, once more with and without assuming isospin symmetry for \BDDs. 
The overall fit quality is very good for all cases, especially when applying isospin symmetry, with which the fit shows excellent compatibility: in this case we obtain
\begin{align}
r_I(D) &= 0.987(36)\,, \qquad
r_I(D^*) = 1.009(42)\,,
\end{align}
again without imposing isospin symmetry in the fit.
We observe that the global fit now reproduces almost exactly the branching fractions and uncertainties obtained in the separate \BD and \BDs fits, but provides also the explicit correlations between all fit parameters. The latter are for instance important in precision determinations of $|V_{cb}|$, but also to be able to include future independent information on the production fractions without having to repeat the present analysis.
We obtain a maximal correlation between the branching ratios of $\sim 30\%$ and further sizable correlations and anti-correlations between the branching fractions and \Rpmz. The input on $f_{\not{B}}$ is asymmetric, which is why we do not include it in the correlation matrix. We checked, however, that all correlations with this parameter are very small in this case, due to the added independent input and the resulting reduced uncertainty.

Our fit combines the information on the production fractions from \BDDs in Table~\ref{tab::BRresults-no-ext} with that from previous global fits \cite{Bernlochner:2023bad,HeavyFlavorAveragingGroupHFLAV:2024ctg}, corresponding to 
$R^{\pm0}_\mathrm{ext} = 1.056(23)$~\cite{Bernlochner:2023bad}. This results in slightly larger central values and smaller uncertainties than in previous global analyses, namely
\begin{align}
    R^{\pm0} &= 1.063(21)\,\, \text{(no isospin assumption in \BDDs), and} \label{eq:Rpm0-global-no-isospin}\\
    &= 1.062(19)\,\, \text{(approx.~isospin for \BDDs)}\,, \label{eq:Rpm0-global-isospin}
\end{align}
see also Table~\ref{tab::Rresults} for \Rpmz values obtained in additional fit configurations.
Both of these values show a deviation of $3\sigma$ from unity, constituting for the first time evidence for isospin violation in $B$ production at $B$ factories. While this value is large compared to naive expectations for isospin violation, due to the nearby threshold, it is actually surprisingly small from a theoretical point of view: it can be accounted for by just considering the relative phase-space factors without any additional enhancement, as already observed in Ref.~\cite{Bernlochner:2023bad}. Importantly, this result demonstrates 
the necessity to account for isospin breaking in $B$ production in the determination of branching ratios and their averages.

\section{Conclusions \label{sec:conclusions}}

In measurements of $B$-meson branching ratios, the corresponding production fractions constitute a main source of uncertainty. Ideally, these would be determined independently from specific $B$ decays, since otherwise the corresponding results cannot be used in the analysis of those decays in a simple manner~\cite{Jung:2015yma,Bernlochner:2023bad}. Given the limited information on $B$-meson production, it is, however, mandatory to use every channel at our disposal. 

In the present work we introduce a new method for the determination of the ratio of production fractions $R^{\pm0}$ based on \BDDs decays.
To that aim we analyze in detail the full set of measurements regarding branching fractions of \BDDs decays. Given the importance of these modes, we extract both the information on their branching fractions and on the production fractions in parallel, providing the correlations in order to avoid double-use of this information in phenomenological analyses.

The results of a comprehensive analysis of the currently available data are given in Table~\ref{tab::BRresults-no-ext}, indicating in particular a value for \Rpmz about $2\sigma$ from unity, similar to other channels, but our method yielding a slightly smaller uncertainty. 
Including the previously available information on the production fractions from other channels~\cite{Bernlochner:2023bad,HeavyFlavorAveragingGroupHFLAV:2024ctg}, we find 
\begin{align}
R^{\pm0} =  1.063(21)\qquad \mathrm{and} \qquad R^{\pm0} =1.062(19)\,,
\end{align}
only the second result assuming approximate isospin symmetry in \BDDs decays, albeit allowing for an enhanced isospin-breaking contribution from electromagnetic interactions. Both values show a significance of $3\sigma$, demonstrating the necessity to include isospin violation in $B$ production consistently in measurements of $B$-meson branching fractions.

In the analysis of the \BDDs branching fractions, we correct several misinterpretations of measurements in the literature and re-analyze several measurements to avoid the d'Agostini bias \cite{DAgostini:1993arp}. We find a number of shifts of both signs, partially cancelling each other, but resulting in a net upwards shift of about $1\sigma$ in both isospin-averaged \BD and \BDs branching fractions, see Table~\ref{tab::BRresults} for details. This result reduces the $V_{cb}$ puzzle, as will be discussed in detail in a forthcoming publication~\cite{Gambino:inpreparation}.

\acknowledgments
We thank Paolo Gambino, Philipp Horak, Marcello Rotondo and Christoph Schwanda for helpful discussions. The work of M.J. is supported in part by the Italian Ministry of University and Research (MUR) under grant PRIN 2022N4W8WR.
S.S.~is supported by the STFC through an Ernest Rutherford Fellowship under reference ST/Z510233/1 and the grant ST/X003167/1.

\appendix

\section{Details on the branching ratio determination \label{app:BRdetails}}

In this appendix we first discuss our treatment of cases with multiple $D$ decay modes and then proceed to provide details for the individual \BD and \BDs measurements.

\subsection{Multiple $D$ final states}\label{app:multiD}

As mentioned in Sec.~\ref{sec:treatment-common}, in many measurements in which several $D$ final states are employed, their relative efficiencies are not published. Given the observations that (i)~the most important $D$ decay mode is usually either $D^0\to K^-\pi^+$ or $D^+\to K^-\pi^+\pi^+$, 
and that (ii)~measurements of the branching ratios of the other decay modes often normalize to the former modes, we treat these cases as follows: 
We write, 
defining $Y_1^0\equiv K^-\pi^+$ and $Y_1^+ \equiv K^+\pi^+\pi^-$,
\begin{align}
&\sum_{i=1}^{N_Y} \epsilon_{D^{0,+}\to Y_i^{0,+}} BR(D^{0,+}\to Y_i) \nn\\
&\qquad =BR(D^{0,+}\to Y_1^{0,+})\left(\epsilon_{D^{0,+}\to Y_1^{0,+}}+\sum_{i=2}^{N_Y}\epsilon_{D^{0,+}\to Y_i} \frac{BR(D^{0,+}\to Y_i^{0,+})}{BR(D^{0,+}\to Y_1^{+,0})}\right)\\
&\qquad =BR(D^{0,+}\to Y_1^{0,+}) \tilde \epsilon_{D^{+,0}\to Y}\,, \label{eq:approx-rescaling}
\end{align}
and only rescale the main branching fraction.
We account for this approximation by subtracting the uncertainty related to the main secondary branching fraction only partially.

\subsection{$B\to D$}

\begin{table*}[t]
\centering{
\resizebox{\textwidth}{!}{
    \begin{tabular}{lllll}\hline\hline
    \multicolumn{5}{c}{\BD}\\\hline
        Analysis    & Data set              & Tagging   & $D^0$ decay modes                             & $D^+$ decay modes\\\hline\hline
        ALEPH'97 \cite{ALEPH:1996dlo}
        & ALEPH $3.9\cdot 10^6$ $Z_{\mathrm{had}}$  &  N/A  & --- & $K^-2\pi^+$\\
        CLEO'98     \cite{CLEO:1998qvx}
        & CLEO 3.16~fb$^{-1}$   & inclusive  & $K^-\pi^+$                                & $K^-2\pi^+$\\
        BaBar'09    \cite{BaBar:2009zxk}
        & BaBar 417~fb$^{-1}$   & tagged& $K^-\pi^+$ + 8 others                           & $K^-2\pi^+$ + 6 others\\
        Belle'15 \cite{Belle:2015pkj}
        & Belle 711~fb$^{-1}$            & tagged    & $K^-\pi^+$ + 12 others                      & $K^-2\pi^+$ + 9 others\\
        Belle-II'25 \cite{Belle-II:2025rna}
        & Belle-II 365~fb$^{-1}$& untagged  & $K^-\pi^+$                                    & $K^-2\pi^+$\\
        \hline\hline
    \end{tabular}
    }
    \caption{Available measurements for \BD.
    \label{tab::B2Dmeasurements}}
}
\end{table*}

We include the results on \BD branching ratios listed in Table~\ref{tab::B2Dmeasurements}. They are discussed in the following, with the exception of the ALEPH result which is discussed further below in Sec.~\ref{sec:ALEPH97} together with the result on \BDs from the same analysis. We proceed in general as described in Sec.~\ref{sec:treatment-common} by defining an effective counting rate or branching ratio.

\subsubsection{CLEO'98 \cite{CLEO:1998qvx}}

 The CLEO measurement~\cite{CLEO:1998qvx} provides only the isospin-averaged rate, \emph{i.e.}, a single observable. The article emphasizes that the two values used as independent inputs in Ref.~\cite{HeavyFlavorAveragingGroupHFLAV:2024ctg} stem from the same rate, 
simply multiplied by different lifetimes. 
    We therefore use $N_\mathrm{eff}$ defined as
    \begin{align}
       N_\mathrm{eff}(\mathrm{CLEO}'98) &= 2 N_{B\bar B} \Gamma(\BD)\times \nn\\
       & \quad \left[f_\pm \tau_{B^-} \mathcal{B}(D^0\to K^-\pi^+) + r_\epsilon f_{00} \tau_{\bar B^0} \mathcal{B}(D^+\to K^-\pi^+\pi^+)\right] \label{eq:Neff-CLEO}\\
        &= 4526\pm 321(\mathrm{stat})\pm320(\mathrm{sys})\,.
    \end{align}
    Here, $r_\epsilon$ is the efficiency ratio between charged and neutral modes, which can be extracted from the given number of events for each mode together with other inputs. 
    Since only the two decay modes $D^0\to K^-\pi^+$ and $D^+\to K^-\pi^+\pi^+$ have been used, the rescaling does not involve the approximation in Eq.~(\ref{eq:approx-rescaling}).

\subsubsection{BaBar'09~\cite{BaBar:2009zxk}}

This measurement uses the inclusive mode \BX as normalization mode, with $E_\ell\geq 0.6\mathrm{GeV}$. We reproduce the values given in this reference only when interpreting the PDG input value for $BR(\BX)$ as that of the neutral $B$ decay. However, the value listed as $BR(\BX)$ by the PDG is actually the isospin-averaged branching fraction, \begin{align}
    BR(\BX)_\mathrm{PDG}=\frac{1}{2}(\tau_{B^0}+\tau_{B^+})\Gamma_\mathrm{iso}(\BX)\,.
\end{align}
We correct for this, which leads to smaller values for the exclusive modes from this analysis. In both cases the neutral branching fraction is multiplied by the lifetime ratio 
\begin{align}
r_\tau=\tau(B^-)/\tau(\bar B^0)    
\end{align}
to obtain the one for the charged mode, \emph{i.e.}, assuming isospin symmetry for the inclusive decay rate. 
    The effective observables are then
    \begin{align}
        r^{N_\mathrm{eff}}_0(\mathrm{BaBar'09}) &= 
        \frac{
            (1+r_\tau)  \mathcal{B}(D^+\to K^-\pi^+\pi^+) \mathcal{B}(\bar B^0\to D^+\ell\bar\nu) 
            }{
            2 \mathcal{B}(\BX)
            } \label{eq:obs-definition-1}\\
        &= (1.92\pm0.09\pm0.08)\times 10^{-2}\,,\\
        r^{N_\mathrm{eff}}_-(\mathrm{BaBar'09}) &= \frac{
            (1+1/r_\tau) \mathcal{B}(D^0\to K^-\pi^+) \mathcal{B}(B^-\to D^0\ell\bar\nu)
            }{
            2 \mathcal{B}(\BX)
            } \label{eq:obs-definition-2}\\
        &= (7.84 \pm0.27\pm0.27)\times 10^{-3}\,, 
    \end{align}
    with a correlation of $50.1\%$, obtained from the given isospin-averaged branching fraction for \BD. As emphasized in Section~\ref{sec:results}, these ratios are independent of the production fractions.

\subsubsection{Belle'15~\cite{Belle:2015pkj}  \label{sec:data-Belle-15}}

The Belle'15~\cite{Belle:2015pkj} analysis uses many $D$ decay modes without providing the relative efficiencies, so we use Eq.~\eqref{eq:approx-rescaling}. We account for d'Agostini bias as described in Sec.~\ref{sec:treatment-common} and illustrated in Fig.~\ref{fig::BtoDBelledAgostini}, which is possible thanks to Belle providing the full correlation matrix for this measurement. This leads to an upwards shift of about $1\sigma$ in the isospin average.

We define the effective yields
    \begin{align}
         N_\mathrm{eff}^{-,0}(\mathrm{Belle}'15) &= 2N_{B\bar B}f_{\pm,0}\tau_{-,0}  \mathcal{B}(D^{0,+}\to K^-\pi^+(\pi^+))\Gamma(B^{-,0}\to D^{0,+}\ell\bar\nu)\,,
    \end{align}
    where we use the rates obtained from the lepton-flavour averages of the Belle data, including their correlations. We obtain   
    \begin{align}
        N_\mathrm{eff}^0(\mathrm{Belle}'15) &= (1.651\pm 0.075)\times 10^6\,, \label{eq:Belle15-neutral}\\ 
        N_\mathrm{eff}^-(\mathrm{Belle}'15) &= (0.795\pm 0.039)\times 10^6\,, \label{eq:Belle15-charged}
    \end{align}
    with a correlation of 73.1\%. 
We note that the maximal correlation between the normalized bins and the total rates is $8\%$,\footnote{This value depends slightly on the (arbitrary) choice of the normalized bin that is removed for being linearly dependent when constructing the correlation matrix.} compared to $63\%$ between the unnormalized bins. This justifies a fit to the total rates, only.

\subsubsection{Belle~II'25~\cite{Belle-II:2025rna}}

Like Ref.~\cite{Belle:2015pkj}, this analysis presents their results as $4\times 10$ bins in $\Delta\Gamma/\Delta w$, providing the full correlation matrix. We account for d'Agostini bias again as described in Sec.~\ref{sec:treatment-common}, extracting the lepton-flavour averages
\begin{align}
\mathcal{B}( B^-\rightarrow D^0 l\bar\nu) &= (2.32 \pm 0.10)\%\,,\\
\mathcal{B}( \bar B^0\rightarrow D^+ l\bar\nu ) &= (2.09 \pm 0.12)\%\,,
\end{align}
with a correlation of $0.6\%$.
The central values of these averages slightly differ
from the corresponding ones given in Ref.~\cite{Belle-II:2025rna} due to our different treatment of the d'Agostini effect.

We define the pseudo observables
\begin{align}
\mathcal{B}_{\mathrm{eff}}^- &= f_{\pm} \mathcal{B}(D^0\rightarrow K^-\pi^+) \mathcal{B}(B^-\rightarrow D^0 l\bar\nu)\,,\\
\mathcal{B}_{\mathrm{eff}}^0 &=  f_{00} \mathcal{B}(D^+\rightarrow K^-\pi^+\pi^+) \mathcal{B}(\bar B^0\rightarrow D^+ l\bar\nu)\,,
\end{align}
for which we obtain
\begin{align}
\mathcal{B}_{\mathrm{eff}}^- &= \left(4.65\pm 0.18\right) \cdot 10^{-4}\,,\\
\mathcal{B}_{\mathrm{eff}}^0 &= \left(9.46\pm 0.49\right)\cdot 10^{-4}\,,
\end{align}
with a correlation of $8.9\%$.

\subsection{$B^-\to D^{*0}\ell\bar \nu$}

We include the results on \BDs branching ratios listed in Table~\ref{tab::B2Dstarmeasurements} as detailed
in the following.

\begin{table*}[t]
\resizebox{\textwidth}{!}{
    \begin{tabular}{lllll}\hline\hline
        \multicolumn{5}{c}{\BDs}\\\hline
        Analysis    & Data set              & Tagging   & $D^{0}$ decay modes                             & $D^{+}$ decay modes \\\hline\hline
        ALEPH'97 \cite{ALEPH:1996dlo}
        & $3.9\cdot 10^6$ $Z_{\mathrm{had}}$  &  N/A & $K^-\pi^+,K_S\pi^+\pi^-,K^-\pi^-2\pi^+$ & \\
        \multirow{2}{*}{OPAL'00~\cite{OPAL:2000hcv}} & \multirow{2}{*}{$\sim 4\cdot 10^6$ $Z_{\mathrm{had}}$} & \multirow{2}{*}{N/A} & $D^0$: inclusive & \\
                                    &                        &     & $D^0$: $K^-\pi^+(\pi^0)$ & \\
        DELPHI~\cite{DELPHI:2001def} & $3\cdot 10^6~Z_{\mathrm{had}}$ & N/A & $D^0$: inclusive & \\
        DELPHI~\cite{DELPHI:2004hkn} & $3.4\cdot 10^6~Z_{\mathrm{had}}$ & N/A & $K^-\pi^+(\pi^0), K^-\pi^-2\pi^+$ & \\
        CLEO'02 \cite{CLEO:2002fch}
        & 3.16~fb$^{-1}$   & untagged & $K^-\pi^+$ & \\
        BaBar'07 \cite{BaBar:2007ddh}
        & 341~fb$^{-1}$   & tagged   & $K^-\pi^+$ + 8 others   & $K^-2\pi^+$ + 8 others \\
        BaBar'07b \cite{BaBar:2007nwi}      
        & 205~fb$^{-1}$   & untagged  & $K^-\pi^+$   &  \\
        BaBar'07c \cite{BaBar:2007cke} & 79~fb$^{-1}$ & untagged & $K^-\pi^+,K^-\pi^+\pi^0, K^-\pi^-2\pi^+$ & \\
            Belle'18    \cite{Belle:2018ezy}
        & $711$~fb$^{-1}$            & untagged  & $K^-\pi^+$                                    & \\
        Belle-II'23b \cite{Belle-II:2023okj}& 189~fb$^{-1}$& untagged  & $K^-\pi^+$   \\
        Belle-II'23c \cite{Belle-II:2023jtw} & 189~fb$^{-1}$ & hadronic tag &  $K^-\pi^+$  \\      \hline\hline
    \end{tabular}}
    \caption{Available measurements for \BDs with their $D$ decay modes.}
    \label{tab::B2Dstarmeasurements}
\end{table*}

\subsubsection{CLEO'02 \cite{CLEO:2002vsd, CLEO:2002fch}}

The main result for this measurement is given again as the isospin-averaged total rate similar to CLEO'98 above, and should therefore again not be used naively as a result for charged and neutral $B$ decays separately. 
However, this measurement also presents results for all four rates $\overline B^{(0,-)}\to D^{*(+,0)}(e,\mu)\bar \nu$, giving explicitly their statistical uncertainties. 
We determine the effective correlation of the statistical error between neutral and charged lepton-flavour averaged decays 
as $-42\%$.
We do not know the correlations of the systematic errors and only take 
the ones resulting from the dependencies on the external inputs into account. We define  
\begin{align}
    & N^{0}_{\mathrm{eff}} = 2 N_{B^0\bar{B}^0} f_{00} \mathcal{B}(D^{*+}\rightarrow D^0 \pi^+)  
    \mathcal{B}(D^0\rightarrow K^-\pi^+ ) \mathcal{B}(B^0\rightarrow D^{*+}l\bar\nu)\qquad\mathrm{and}\\
    & N^{+}_{\mathrm{eff}} = 2 N_{B^+B^-} f_{\pm} \mathcal{B}(D^{*0}\rightarrow D^0\pi^0)\mathcal{B}(\pi^0\rightarrow \gamma\gamma) 
    \mathcal{B}(D^0\rightarrow K^-\pi^+ ) \mathcal{B}(B^-\rightarrow D^{*0}l\bar \nu)\,,
    \end{align}
    for which we obtain:
    \begin{align}
    N^{0}_{\mathrm{eff}}    = 5055\pm 253 \pm 292\,\qquad\mathrm{and}\qquad
    N^{+}_{\mathrm{eff}}    = 5060\pm 329\pm 367\,,
    \end{align}   
    with a total correlation of $-18.3\%$.

\subsubsection{BaBar'07a~\cite{BaBar:2007ddh} }

This analysis, similarly to BaBar'09, uses the inclusive decay \BX as normalization mode. We reproduce the result for the neutral branching fraction by again interpreting the PDG \BX branching fraction as that of the neutral decay, including an extrapolation factor to relate the total rate given in the PDG to the one with a lepton-energy cut measured in the analysis; however, we reproduce the charged branching fraction by interpreting the \emph{same} branching fraction as that of the charged decay, \emph{and} removing the extrapolation factor.\footnote{While it is a choice to include the extrapolation factor in the efficiency or not, it seems extremely unlikely that this choice has on purpose been made differently for the charged and neutral decay, so we consider this a mistake in the analysis; we consider the choice also made in the later analysis Ref.~\cite{BaBar:2009zxk} the correct one.}
Correcting for these factors, we obtain for
the ratios $r_{0,-}^\mathrm{eff}$, defined analogously to the $B\to D$ case,
\begin{align}
r_0^{N_{\mathrm{eff}}}(\text{BaBar'07}) &= \frac{1 +r_{\tau}}{2} \frac{ \mathcal{B}(\bar{B}^0\rightarrow D^{*+} l\bar\nu) \mathcal{B}(D^{*+}\rightarrow D^0\pi^+) \mathcal{B}(D^0\rightarrow K^-\pi^+) 
        }{ 
            \mathcal{B}(\BX) 
            } \\ 
            &= 0.0131\pm 0.0004 \pm 0.0005\,, \\ \nn\\
 r_-^{(N_{\mathrm{eff}})}(\text{BaBar'07}) &= \frac{1 + 1/r_{\tau}}{2} \frac{ \mathcal{B}(B^-\rightarrow D^{*0}l\bar\nu) \mathcal{B}(D^0\rightarrow K^-\pi^+)
            }{
                \mathcal{B}(\BX)
                } \\ 
            &= 0.0196\pm 0.0005 \pm 0.0009\,.
\end{align}

\subsubsection{BaBar'07b~\cite{BaBar:2007nwi}}

The BaBar'07b analysis obtains the branching ratio as a fit to the CLN parametrization, hence the comments from Sec.~\ref{sec:treatment-common} apply. They quote the derivatives $\partial\mathcal{B}/\partial R_1(1)$ and  $\partial\mathcal{B}/\partial R_2(1)$. These allow to rescale the branching ratio to current values of $R_1(1)$ and $R_2(1)$.
    We define the pseudo-observable: 
    \begin{align}
     N_{\mathrm{eff}}(\text{BaBar'07b}) &= 2 N_{B\bar{B}} f_{\pm} 
            \mathcal{B}(B^-\rightarrow D^{*0}l\bar \nu) \mathcal{B}(D^{*0}\rightarrow D^0\pi^0)  \mathcal{B}(D^0\rightarrow K\pi) 
        \mathcal{B}(\pi^0\rightarrow \gamma\gamma) \nn\\
    &= 299520 \pm 4252 \pm 14736\,.
    \end{align}

\subsection{$\bar B^0\to D^{*+}\ell\bar \nu$}

\subsubsection{ALEPH'97~\cite{ALEPH:1996dlo} \label{sec:ALEPH97}}

    ALEPH'97~\cite{ALEPH:1996dlo} reconstructs several exclusive $D$ final states, but does not provide information on the individual channels, so we have to rely on the approximate procedure outlined above. In addition to $\bar B^0\to D^{*+}\ell\bar\nu$ they also provide a measurement of a combination of $\bar B^0\to D^+\ell\bar\nu$ and $\bar B^0\to D^{*+}(\to D^+\gamma/\pi^0)\ell\bar\nu$ decays, which is used to constrain the $D$ mode. 
    Consequently, we define the following pseudo-observables:
    \begin{align}
     \mathcal{B}_{\mathrm{eff}}^{\mathrm{ALEPH}}(\bar{B}^0\rightarrow D^{*+}l\bar\nu) &= 
        R_b f_{B^0} \mathcal{B}(D^{*+}\rightarrow D^0\pi^+) \mathcal{B}(D^0\rightarrow K^-\pi^+)    
        \mathcal{B}(\bar{B}^0\rightarrow D^{*+}l\bar\nu) \\
        &=  (1.210 \pm 0.057 \pm 0.072 )\cdot 10^{-4}\,,  \\
      \mathcal{B}_{\mathrm{eff}}^{\mathrm{ALEPH}}(\bar{B}^0\rightarrow D^+ l\bar\nu) &= R_b f_{B^0} \mathcal{B}(D^+\rightarrow K^-\pi^+\pi^+) \Sigma \mathcal{B}^{\mathrm{ALEPH}}\\
            &= ( 2.33 \pm 0.15 \pm 0.34 ) \cdot 10^{-4}\,, 
    \end{align}
    where 
    \begin{align}
    &\Sigma \mathcal{B}^{\mathrm{ALEPH}} \equiv 
        \mathcal{B}(\bar{B}^0\rightarrow D^+l\bar\nu) + (1 - \mathcal{B}(D^{*+}\rightarrow D^0 \pi^+)) \varepsilon_r^\mathrm{ALEPH} \mathcal{B}(\bar{B}^0\rightarrow D^{*+} l\bar\nu )\,.
    \end{align}
  The relative efficiency $\varepsilon_r^\mathrm{ALEPH}$ is determined  from 
    \begin{align}
    N(K^-\pi^+\pi^+)_{D^{*+}} &=  N_{B^0} \mathcal{B}( \bar{B}^0\rightarrow D^{*+} l\bar\nu) \left(1 - \mathcal{B}(D^{*+} \rightarrow D^0\pi^+ ) \right)\times \nn\\ &\quad                \mathcal{B}( D^+\rightarrow K^-\pi^+\pi^+ ) 
                \varepsilon_{D^*}^{\mathrm{ALEPH}}\,, \\
    N(K^-\pi^+\pi^+)_{D^+} &= N_{B^0} \mathcal{B}(\bar{B}^0\rightarrow D^+ l\bar\nu) \mathcal{B}(D^+\rightarrow K^-\pi^+\pi^+) \varepsilon_{D}^{\mathrm{ALEPH}}\,,
    \end{align}
    as
    \begin{align}
    \varepsilon_{r}^{\mathrm{ALEPH}} \equiv \frac{  
        \varepsilon_{D^*}^{\mathrm{ALEPH}}
    }{
        \varepsilon_{D}^{\mathrm{ALEPH}}
    } &= 
    \frac{
        N(K^-\pi^+\pi^+)_{D^{*+}} \mathcal{B}(\bar{B}^0\rightarrow D^+l\bar\nu)
        }{
        N(K^-\pi^+\pi^+)_{D^+} 
        \left( 1 - \mathcal{B}(D^{*+}\rightarrow D^0\pi^+) \right)
        \mathcal{B}(\bar{B}^0\rightarrow D^{*+}l\bar\nu) 
        }\,,
    \end{align}
using the numerical values from within the analysis.

\subsubsection{OPAL'00~\cite{OPAL:2000hcv}}

OPAL employs two strategies for the measurement of $\bar B^0\to D^*l\bar \nu$ decays: exclusive reconstruction, similar to the measurements discussed above, and \emph{inclusive} reconstruction with respect to the $D$ decay, using only the slow pion from the $D^*\to D\pi$ decay.
They provide sufficient information to rescale all $D$ branching fractions involved and to determine the correlation between the inclusive and exclusive measurements. We introduce the corresponding pseudo-observables 
    \begin{align}
     \mathcal{B}_{\mathrm{eff}}^{\mathrm{OPAL,incl}} &\equiv 
            R_b f_{B^0} \mathcal{B}(D^{*+}\rightarrow D^0 \pi^+) \mathcal{B}(\bar B^0\rightarrow D^{*+} l\bar\nu)_{\mathrm{incl}} \\
            &= (3.48\pm 0.16 \pm 0.34)\cdot 10^{-3}\\
    \mathcal{B}_{\mathrm{eff}}^{\mathrm{OPAL,excl}} &\equiv 
        R_b f_{B^0} \mathcal{B}(D^{*+} \rightarrow D^0 \pi^+) \left(\mathcal{B}(D^0\rightarrow K^-\pi^+) + r_{\varepsilon}^{\mathrm{OPAL}} 
            \mathcal{B}(D^0\rightarrow K^-\pi^+\pi^0 ) \right) \times\nn\\
          &\quad\mathcal{B}(\bar B^0\rightarrow D^{*+}l\bar\nu)_{\mathrm{excl}} \nn\\
        & = (2.89\pm 0.11 \pm 0.19) \cdot 10^{-4}\,, 
    \end{align}
    with a correlation of $31.5\%$ between the measurements.

\subsubsection{DELPHI'01~\cite{DELPHI:2001def} and DELPHI'04~\cite{DELPHI:2004hkn}}
    
The DELPHI collaboration performed again both an inclusive \cite{DELPHI:2001def} and an exclusive \cite{DELPHI:2004hkn} measurement. The latter analysis provides an update of and combination with the former, which allows to obtain the approximate correlation between the two measurements. Also the DELPHI collaboration provides sufficient information to update the involved $D$ meson branching fractions. 
For the exclusive measurement we define the effective observable
\begin{align}
\mathcal{B}_{\mathrm{eff}}^{\mathrm{excl}}(\mathrm{DELPHI}) &=
R_b f_{B^0} \mathcal{B}(D^{*+}\rightarrow D^0\pi^+) 
\mathcal{B}(\bar B^0\rightarrow D^{*-}l\bar \nu) \times \nn\\
&\hspace{-1cm}\left( \mathcal{B}(D^0\rightarrow K\pi) + 
        \frac{
            \varepsilon_{D^0K\pi\pi\pi}
            }{
            \varepsilon_{D^0K\pi}
            }
        \mathcal{B}(D^0\rightarrow K\pi\pi\pi) + \frac{
            \varepsilon_{D^0K\pi\pi}
            }{
            \varepsilon_{D^0K\pi}
            }
        \mathcal{B}(D^0\rightarrow K\pi\pi)
  \right) \\
  &=  (4.49\pm0.35)\cdot 10^{-4}\,. 
\end{align}
For the inclusive measurement we employ the same pseudo-observable as for OPAL: 
\begin{align}
\mathcal{B}_{\mathrm{eff}}^{\mathrm{incl}}(\mathrm{DELPHI}) &= R_b f_{B^0} \mathcal{B}(D^{*+}\rightarrow D^0\pi^+) \mathcal{B}(\bar B^0\rightarrow D^{*-}l\bar \nu)\\
&=(2.96\pm0.23)\cdot 10^{-3}\,. 
\end{align}
We obtain a correlation of $53.2\%$ between the two observables.

\subsubsection{BaBar'07c~\cite{BaBar:2007cke}}

The untagged BaBar measurement~\cite{BaBar:2007cke} performs a CLN fit to three single-differential distributions and gives a final branching fraction obtained from integrating the theoretical expression using the fitted parameters. It is thereby sensitive to significant d'Agostini bias, as discussed in Sec.~\ref{sec:treatment-common}.
Importantly, this analysis provides sufficient information to crosscheck our approximate procedure of rescaling only the main $D$ branching fractions, see Eq.~(\ref{eq:approx-rescaling}): the presented information allows to obtain the relative efficiencies for the three decay modes, such that the effect from updating all three $D^0$ decay modes used in the analysis can be compared to applying our procedure of rescaling only the main mode. We find agreement within uncertainties,
despite sizable changes in all branching fractions involved, confirming our approach. Nevertheless, since in this case sufficient information is available, we 
use the observable
    \begin{align}
         N_\mathrm{eff}^{0} &= 2N_{B\bar B} f_{00} \mathcal{B}(\bar B^0\to D^{*+}\ell\bar \nu) \mathcal{B}(D^{*+}\to D^0\pi^+) \mathcal{B}_\mathrm{eff}(D^0)\\
        &=(2.35\pm 0.16) \cdot 10^5\,, 
    \end{align}
    with
    \begin{align}
         \mathcal{B}_\mathrm{eff}(D^0) &= \mathcal{B}(D^0\to K^-\pi^+) + \epsilon^{K^-\pi^+\pi^+\pi^-}_\mathrm{rel} BR(D^0\to K^-\pi^+\pi^+\pi^-) + \nonumber\\
        &\quad \epsilon^{K^-\pi^+\pi^0}_\mathrm{rel} BR(D^0\to K^-\pi^+\pi^0)\,.
    \end{align}

\subsubsection{Belle'18 \cite{Belle:2018ezy}}

This analysis presents an untagged measurement of 4 single-differential rates $d\Gamma/dx$, with $x\in\{w,\cos\theta_\ell,\cos\theta,\chi\}$, in 10 bins each, in the two modes $\bar B^0\to D^{*+}\ell\bar\nu$, with $\ell=e,\mu$. 
The angular distribution exhibits a tension with LFU at the level of $\sim 4\sigma$, independent of the treatment of the $e-\mu$ correlations~\cite{Bobeth:2021lya}.
We interpret this tension as underestimated systematic uncertainties, consistent with more recent studies~\cite{Belle:2023bwv,Belle-II:2023svm}, and discard the angular distributions from this study.
We retain the information on the differential distribution in $w$, however, in the form of normalized bins $\hat N_i=N_i/(\sum_j N_j)$, in order to avoid the d'Agostini bias. 
We obtain
\begin{align}
N^{0}_{\mathrm{eff}} &\equiv 
2 f_{00} N_{B\bar{B}} \mathcal{B}(D^{*+}\rightarrow D^0\pi^+) \mathcal{B}(D^0\rightarrow K^- \pi^+)  \mathcal{B}(\bar B^0\to D^{*+}\ell\bar\nu)\\
&= (9.99\pm 0.24)\cdot 10^5.
\end{align}

\subsubsection{Belle-II'23b~\cite{Belle-II:2023okj}}

This analysis presents measurements of 4 single-differential rates $d\Gamma/dx$, with $x\in\{w,\cos\theta_\ell,\cos\theta,\chi\}$, in 8 bins for the $\cos\theta_\ell$ distribution and 10 bins for each of the others, in the two modes $\bar B^0\to D^{*+}\ell\bar\nu$, with $\ell=e,\mu$, based on 189~fb$^{-1}$ of Belle~II data.
The total rate is calculated by simply summing the bin contents of any of the distributions, which should be unaffected from d'Agostini bias.
We obtain
\begin{align}
N^{0}_{\mathrm{eff}} &\equiv 2  f_{00} N_{B\bar{B}} \mathcal{B}(D^{*+}\rightarrow D^0\pi^+) \mathcal{B}(D^0\rightarrow K^- \pi^+)  \mathcal{B}(\bar B^0\to D^{*+}\ell\bar\nu)\\
    &= (2.52 \pm 0.08)\cdot 10^5\,.
\end{align}

\subsubsection{Belle-II'23c~\cite{Belle-II:2023jtw}}

This hadronically tagged analysis uses CLN throughout, hence the comments in Sec.~\ref{sec:treatment-common} apply. The study is not yet published, but also included in Ref.~\cite{HeavyFlavorAveragingGroupHFLAV:2024ctg}, so we include it as well to facilitate comparisons.
We obtain the corresponding  pseudo-observable
\begin{align}
N^{0}_{\mathrm{eff}} &\equiv 
2  f_{00} N_{B\bar{B}} \mathcal{B}(D^{*+}\rightarrow D^0\pi^+) \mathcal{B}(D^0\rightarrow K^- \pi^+)  \mathcal{B}(\bar B^0\to D^{*+}\ell\bar\nu)\\
    &= (2.70\pm 0.21)\cdot 10^5\,.
\end{align}

\bibliography{draft.bib}
\bibliographystyle{apsrev4-1}

\end{document}